\documentclass[aps,preprint,prd,nofootinbib]{revtex4-1}
\pdfoutput=1
\usepackage{graphicx}
\usepackage{psfrag}
\usepackage{float}
\usepackage{subfigure}
\usepackage{array}
\usepackage{graphicx}
\usepackage{amsmath}
\usepackage{amssymb}
\usepackage{mathrsfs}
\usepackage{bm}
\usepackage{slashed}
\usepackage{threeparttable}
\usepackage{multirow}
\usepackage[colorlinks, linkcolor=black, anchorcolor=black, citecolor=black]{hyperref}
\usepackage[amsmath,thmmarks,hyperref]{ntheorem}
\usepackage{soul}
\allowdisplaybreaks[4]

\begin{document}

\title{The light invisible boson in FCNC decays of $B$ and $B_c$ mesons}
\author{Geng Li$^1$\footnote{karlisle@hit.edu.cn}, Tianhong Wang$^1$\footnote{thwang@hit.edu.cn (Corresponding author)}, Jing-Bo Zhang$^1$\footnote{jinux@hit.edu.cn} and Guo-Li Wang$^{1, 2, 3}$\footnote{gl\_wang@hit.edu.cn}\\}
\address{$^1$School of Physics, Harbin Institute of Technology, Harbin, 150001, China\\
$^2$Department of Physics, Hebei University, Baoding 071002, China\\
$^3$Hebei Key Laboratory of High-precision Computation and Application of Quantum Field Theory, Baoding
071002, China}

\baselineskip=20pt

\begin{abstract}
In this paper, we study the FCNC decay processes of $B$ and $B_c$ meson, in which one invisible particle is emitted. Both the spin-0 and spin-1 cases are considered. The model-independent effective Lagrangian is introduced to describe the coupling between the light invisible boson and quarks. The constraints of the coupling coefficients are extracted by experimental upper limits of the missing energy in $B$ meson decays. The bounds are used to predict the upper limits of branching fractions of corresponding $B_c$ decays, which are of the order of $10^{-6}$ or $10^{-5}$ when final meson is pseudoscalar or vector, respectively. The maximum branch ratios are achieved when $m_\chi\approx3.5- 4$~GeV, where $m_\chi$ is the mass of the invisible particle. 
\end{abstract}

\maketitle

\section{Introduction}
Dark matter (DM) played an important role in the evolution of the universe. The freeze-out mechanism~\cite{Bernstein:1985th, Srednicki:1988ce} considered dark matter candidates as thermal relic from the local thermodynamic equilibrium of early universe~\cite{Izaguirre:2015yja}. Their annihilation cross sections are bounded by the observed dark matter relic abundance $\Omega_c h^2 = 0.1131\pm 0.0034$~\cite{Bertone:2004pz, Komatsu:2008hk}. Interestingly, this limitation of interaction intensity happens to be on the same order of magnitude as that of weak interaction, which makes the weakly-interacting massive particle (WIMP) to be one of the most promising dark matter candidates. Currently, the direct and indirect DM detections~\cite{Akerib:2016vxi, Cui:2017nnn, Aprile:2018dbl} get null results and set much stricter constraints on the parameter space for the WIMP with mass larger than several GeV. It provides a motivation for the study of light dark matter candidates through high-energy colliders, for example, CODEX-b at the LHCb experiment aimed to probe for GeV-scale long-lived particles~\cite{Gligorov:2017nwh}. The Lee-Winberg~\cite{PhysRevLett.39.165} limit which sets the lower bound of the WIMP mass to a few GeV is a model-dependent result. This constraint can be relaxed with different models or proper parameters selection. It makes lower mass WIMP be possible, for example, the MeV-scale light dark matter (LDM) is proposed~\cite{Pospelov:2007mp,Hooper:2008im} to explain the unexpected emission of 511 keV photons from the galaxy center. The feebly interacting massive particle (FIMP) is another DM candidate which comes from an alternative scenario of the freeze-in mechanism~\cite{McDonald:2001vt, Hall:2009bx,Bernal:2017kxu}. Within the freeze-in scenario, the DM is never in thermal equilibrium with the SM and is gradually produced from scattering or decay of the Standard Model (SM) particles. It allows much weaker interaction between the SM particles and DM. 

High-energy collider searches might be able to detect dark matter particles produced in collisions through their invisible (“missing”)
energy and momentum, which do not match SM neutrino prediction. This motivates us to study whether DM interactions could help to explain the anomalies. So far, these experiments provide mostly just upper limits on the interaction strength between DM and the SM. The BaBar and Belle~\cite{delAmoSanchez:2010bk, Aubert:2008am, Chen:2007zk,Grygier:2017tzo,Lai:2016uvj} known as B-factories produce large numbers of $B$ mesons, allowing to study their various decay channels precisely, which has revealed tentative anomalies with respect to SM predictions. New models involved invisible particles have been extensively studied in the flavor-changing neutral current (FCNC) processes~\cite{Bird:2004ts,Bird:2006jd,Badin:2010uh,Gninenko:2015mea,Barducci:2018rlx,Kamenik:2011vy,Bertuzzo:2017lwt}. While previous studies most focus on $B$ meson instead of $B_c$ meson. The $B_c$ meson has been massively produced and measured by the CDF~\cite{Aaltonen:2016dra}, ATLAS~\cite{Burdin:2016rzf}, CMS~\cite{Berezhnoy:2019yei}, and LHCb~\cite{Aaij:2019ths}  experiments. The production rate of $B_c$ meson on the LHCb collaboration is close to 3.7 per mille of that of the $B$ mesons~\cite{Aaij:2019ths}. The $B_c$ events are of the order of $10^{10}$ per year. As the luminosity of the LHC increases significantly, much more $B_c$ events will be generated in the near future, which provides a new possibility to discover dark matter candidates. 

Except for photons, the SM bosons cannot exist stably for a long time. In models, the invisible boson can either be the stable relics in previous Universe or a mediator between the SM and dark sector. Vector dark matter (VDM)~\cite{Pospelov:2008jk, Redondo:2008ec, Bjorken:2009mm} candidates are usually introduced through Abelian or non-Abelian extended gauge group. In order to make VDM itself a candidate for dark matter, additional symmetries are often requested to maintain its stability~\cite{DiazCruz:2010dc, Baek:2012se}. A well-know invisible vector model is the dark photon~\cite{Fabbrichesi:2020wbt}. A very light massive dark photon could be a dark matter candidate, while in other cases, dark photon appears as a mediator. One of spin-0 hidden boson candidates is the axion-like pseudoscalar particle.  Axion was introduced in order to explain the strong-CP problem~\cite{Peccei:1977hh, PhysRevLett.40.223, Wilczek:1977pj}. Axion-like dark matter (ALDM) models~\cite{Batell:2009jf, Aditya:2012ay, Izaguirre:2016dfi} usually introduce a general dimension-five Lagrangian which consists of scalar and vector current to describe the coupling between SM fermions and ALDMs. Scalar dark matter candidates can be achieved in minimal extensions of the SM~\cite{OConnell:2006rsp, Patt:2006fw}, in which the hidden scalar can mix with the Higgs boson~\cite{Krnjaic:2015mbs, Winkler:2018qyg, Filimonova:2019tuy, Kachanovich:2020yhi}. If the scalar further decays into double leptons $l\bar l$, it is possible to observe this signal in the experiments. If it decays into two invisible fermions $\bar \chi \chi$, the scalar is a mediator between the SM and the dark sector.

In this paper, we focus on the light invisible bosonic particle (both scalar and vector) which is emitted in FCNC decays of $B$ and $B_c$ meson. We introduce a general dimension-5 effective Lagrangian which includes coupling strength of quarks and an invisible boson. The Wilson coefficients are extracted from the experimental results of the $B$ meson decays with missing energy, which are used to predict the upper limits of
the branching fractions of the similar decay processes of $B_c$ meson. 

The paper is organized as follows: In Sec.~II, we study the decay processes of $B$ and $B_c$ mesons with single invisible scalar ($\chi=S$) production. In Sec.~III, we study the single invisible vector ($\chi=V$) generated case. Finally, we draw the conclusion in Sec.~IV.

\section{Light invisible scalar}

\begin{figure}[h]
	\centering
	\includegraphics[width=0.4\textwidth]{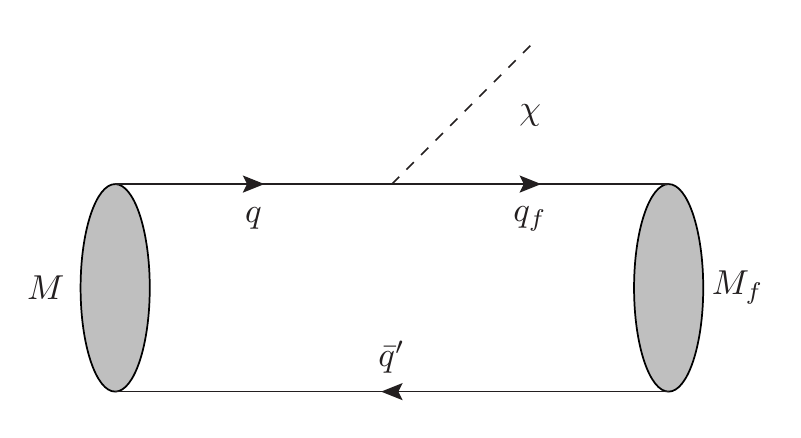}
	\caption{Feynman diagram of decay channels involving invisible particles,}
	\label{Feyn01}
\end{figure}

The experimental upper limits of $B$ meson FCNC decays with missing energy from Belle Collaboration and SM predictions are listed in Table.~\ref{tab1}. 
\begin{table}[h]
	\setlength{\tabcolsep}{0.5cm}
	\caption{The branching ratios (in units of $10^{-6}$) of $B$ meson decays involving missing energy.}
	\centering
	\begin{tabular*}{\textwidth}{@{}@{\extracolsep{\fill}}ccc}
		\hline\hline
		Experimental bound~\cite{Chen:2007zk,Grygier:2017tzo,Lai:2016uvj}&SM prediction~\cite{Kamenik:2009kc,Jeon:2006nq,Altmannshofer:2009ma,Bartsch:2009qp}&Invisible particles bound\\
		\hline
		${\rm BR}(B^\pm\to K^\pm\slashed E)<14$& ${\rm BR}(B^\pm\to K^\pm\nu \bar{\nu})=5.1 \pm 0.8$  & ${\rm BR}(B^\pm\to K^\pm\chi)<9.7$	\\
		${\rm BR}(B^\pm\to \pi^\pm\slashed E)<14$& ${\rm BR}(B^\pm\to \pi^\pm\nu \bar{\nu})=9.7 \pm 2.1$  & ${\rm BR}(B^\pm\to \pi^\pm\chi)<6.4$	\\
		${\rm BR}(B^\pm\to K^{*\pm}\slashed E)<61$& ${\rm BR}(B^\pm\to K^{*\pm}\nu \bar{\nu})=8.4 \pm 1.4$  & ${\rm BR}(B^\pm\to K^{*\pm}\chi)<54$	\\
		${\rm BR}(B^\pm\to \rho^\pm\slashed E)<30$&${\rm BR}(B^\pm\to \rho^\pm \nu \bar{\nu})=0.49^{+0.61}_{-0.38}$  & ${\rm BR}(B^\pm\to \rho^\pm \chi)<30$	\\
		\hline\hline
		\label{tab1}
	\end{tabular*}
\end{table}
It can be seen that the theoretical prediction is smaller than the experimental value, which leaves room for contributions from new physics~\cite{Grygier:2017tzo}. We assume that a hidden boson $\chi$ produced in these processes carries away part of energy. The  Feynman diagram is presented as in Fig.~\ref{Feyn01}, 
where $q$, $q_{_f}$, and $\bar q^\prime$ represent the quark and antiquark, $M$ and $M_f$ are the masses of the initial and final mesons, respectively. When $\chi=S$, we introduce a dimension-5 model-independent effective Lagrangian to describe the vertex which represents the coupling between SM fermions and the hidden scalar, 
\begin{equation}
	\begin{aligned}
		\mathcal L_{scalar}=m_{_S} g_{_{S1}} (\bar q_{_f} q) S +m_{_S} g_{_{S2}} (\bar q_{_f} \gamma^{5} q) S +g_{_{S3}} (\bar q_{_f}\gamma_{\mu}q)(i\partial^\mu S) + g_{_{S4}} (\bar q_{_f} \gamma_{\mu}\gamma^{5}q)(i\partial^\mu S),
		\label{eq1}
	\end{aligned}
\end{equation}
where $g_{_{Si}}$s are phenomenological coupling constants. The operators $(\bar q_{_f} q) S$ and $(\bar q_{_f} \gamma^{5} q) S$ break ${\rm SU(2)}_L$ symmetry, as (pseudo)scalar currents are necessarily involving quarks with opposite chirality. If one starts from an effective Lagrangian which respects the SM gauge symmetry, these operators could be suppressed severely. For example, Ref.~\cite{Kamenik:2011vy} included operators like $(H^\dagger\bar q_{_f} q) S$ and $(H^\dagger\bar q_{_f} \gamma^5 q) S$ by considering the electroweak symmetry breaking. These coefficients are suppressed by an additional factor $v/\Lambda$ with $v$ being the vacuum expectation value of Higgs field and $\Lambda$ being the new physics scale (usually considered to be in TeV). In this research we are interested that to what extent the experimental data will constrain these coefficients. The operators are just introduced phenomenologically instead of starting from gauge symmetry. 

Similar processes were discussed in some previous papers, for example, Ref.~\cite{Krnjaic:2015mbs, Winkler:2018qyg, Filimonova:2019tuy, Kachanovich:2020yhi} considered the hidden scalar can mix with Higgs boson and introduced a coupling Lagrangian with mixing angle $\theta$. The experimental limits of $B$ and $K$ meson decays are used to set bounds for $\theta$. Ref.~\cite{Pospelov:2008jk} discussed about constraints of keV-scale bosonic DM candidates. In this work, we study the bosonic DM candidate with mass of several GeV, and set upper limits for the branching ratios of $B_c$ meson decays with the emission of the hidden boson. 

\subsection{$0^-\to 0^-$ meson decay processes}
According to the Feynman diagram and the effective Lagrangian, the amplitude of $0^-\to 0^-$ meson decay can be written as
\begin{equation}
	\begin{aligned}
		\langle M_fS|\mathcal L_{scalar}|M \rangle & =m_{_S}g_{_{S1}}\langle M_f^-|(\bar q_{_f} q)|M^-\rangle+g_{_{S3}}	\langle M_f^-|(\bar q_{_f}\gamma_\mu q) |M^-\rangle P_{_S}^\mu\\
		&=g_{_{S1}} \mathcal {T}_1+g_{_{S3}} \mathcal {T}_3,
		\label{eq1.2}
	\end{aligned}
\end{equation}
where $\mathcal T_i$s are amplitudes other than the effective coupling coefficients. $P_{_S}$ is the  momenta of the invisible scalar. As the Lagrangian is sum of several operators, the partial width can be written as
\begin{equation}
	\begin{aligned}
		\Gamma =\int {dPS_2 \big(\sum_{j} g_{_{Sj}}\mathcal{T}_j\big)^{\dagger} \big(\sum_{i} g^{ }_{_{Si}} \mathcal{T}_i \big) }=\sum_{ij}g_{_{Sj}}g_{_{Si}}\widetilde\Gamma_{ij}, 
		\label{eq2}
	\end{aligned}
\end{equation}
Here we have defined $\widetilde\Gamma_{1(3)}=\int dPS_3 |\mathcal{T}_{1(3)}|^2$, which are independent of the coefficients. When the final meson is a pseudoscalar, by finishing the two-body phase space integral, we get the decay width. 
\begin{equation}
	\begin{aligned}
		\Gamma(M\to M_fS)=&\frac{1}{16\pi M^3 } \lambda^{1/2}(M^2,M_f^2,m_{_S}^2) \bigg\{ m_{_S}^2 g_{_{S1}}^2 \langle M_f^-|(\bar q_{_f} q)|M^-\rangle^*\langle M_f^-|(\bar q_{_f} q)|M^-\rangle \\
		&+ g_{_{S3}}^2 \langle M_f^-|(\bar q_{_f} \gamma_\nu q)|M^-\rangle^*\langle M_f^-|(\bar q_{_f} \gamma_\mu q)|M^-\rangle P_{_S}^\nu P_{_S}^\mu \\
		&+m_{_S} g_{_{S1}} g_{_{S3}}^* \langle M_f^-|(\bar q_{_f} \gamma_\nu q)|M^-\rangle^*\langle M_f^-|(\bar q_{_f}  q)|M^-\rangle P_{_S}^\nu +h.c. \bigg\}, 
		\label{eq3}
	\end{aligned}
\end{equation}
where the K${\rm \ddot a}$llen function $\lambda(x, y, z)= x^2 + y^2 +z^2 -2xy-2xz -2yz$ is used. The hadronic transition matrix elements can be expressed as
\begin{equation}
	\begin{aligned}
		\langle M_f^-|(\bar q_{_f} q)|M^-\rangle&\simeq \frac{(P-P_f)^\mu}{m_q-m_{q_{_f}}}\langle M_f^-|(\bar q_{_f}\gamma_\mu q) |M^-\rangle= \frac{M^2-M_{f}^2}{m_q-m_{q_{_f}}}f_0 (s), \\
		\langle M_f^-|(\bar q_{_f}\gamma_\mu q) |M^-\rangle 
		&= (P+P_f)_{\mu}f_+ (s)+(P-P_f)_{\mu}\frac{M^2-M_{f}^2}{s} \big[f_0 (s)-f_+ (s)\big],
		\label{eq4}
	\end{aligned}
\end{equation}
where $s=(P-P_f)^2=m_{_S}^2$; $f_0$ and $f_+$ are form factors; $m_q$ and $m_{q_f}$ are the masses of initial and final quarks, respectively. It is worth to mention that one of the form factors in Eq.~(\ref{eq4}) is divergent when $m_{_S}=0$, however, the final results are smooth and convergent when $m_{_S}\to 0$. The hadronic matrix element with pseudoscalar current $\langle M_f^-|(\bar q_{_f} \gamma^5 q)|M^-\rangle$ and axial vector current $\langle M_f^-|(\bar q_{_f} \gamma_\mu\gamma^5 q)|M^-\rangle$ are zero for the $0^-\to0^-$ processes. When we calculate the hadronic matrix elements of $B$ meson decays, the LCSR method are adopted to write the form factors~\cite{Ball:2004ye}. One can see more details of the selection of parameters in our previous work~\cite{Li:2018hgu, Li:2020dpc}. The instantaneous Bethe-Salpeter (BS) method ~\cite{Kim:2003ny, Wang:2005qx} which is more suitable for heavy to heavy meson decays is used in $B_c$ meson decay processes. In Mandelstam formalism, the hadronic transition matrix element is written as
\begin{equation}
	\begin{aligned}
		\langle h^-|\bar q_1\Gamma^\xi b|B_c^-\rangle 
		&= \int\frac{d^3 q}{(2\pi)^3} {\rm Tr}\left[\frac{\slashed P}{M}\overline\varphi_{P_f}^{++}(\vec q_{f})\Gamma^\xi\varphi_P^{++}(\vec q)\right],
		\label{eq5}
	\end{aligned}
\end{equation}
where $\Gamma^\xi=1,\gamma^5,\gamma^\mu,\gamma^\mu\gamma^5$ or $\sigma^{\mu\nu}$; $\varphi^{++}_P$ and $\varphi^{++}_{P_f}$ are the wave functions of the initial and final mesons, respectively; $P$ and ${P_f}$ are the momenta of the initial and final mesons, respectively; $\vec q$ and $\vec q_{_f}$ are the relative momenta of the quark and antiquark in the initial and final meson, respectively. 

\begin{figure}[h]
	\centering
	\subfigure[$B^-\rightarrow K^-S$]{		\label{width-S14a}
		\includegraphics[width=0.45\textwidth]{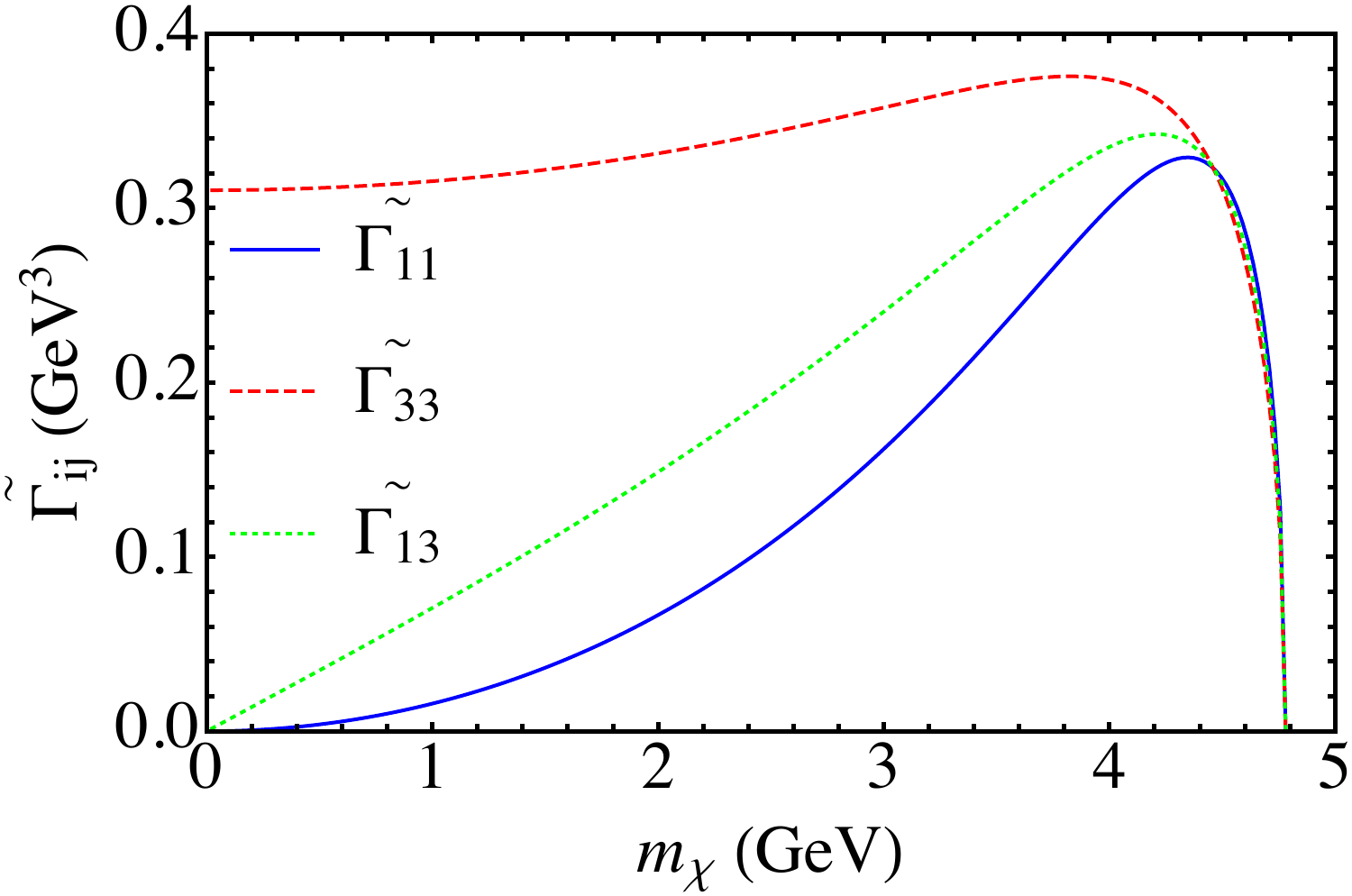}}
	\hspace{2em}
	\subfigure[$B^-\rightarrow \pi^-S$]{		\label{width-S14b}
		\includegraphics[width=0.45\textwidth]{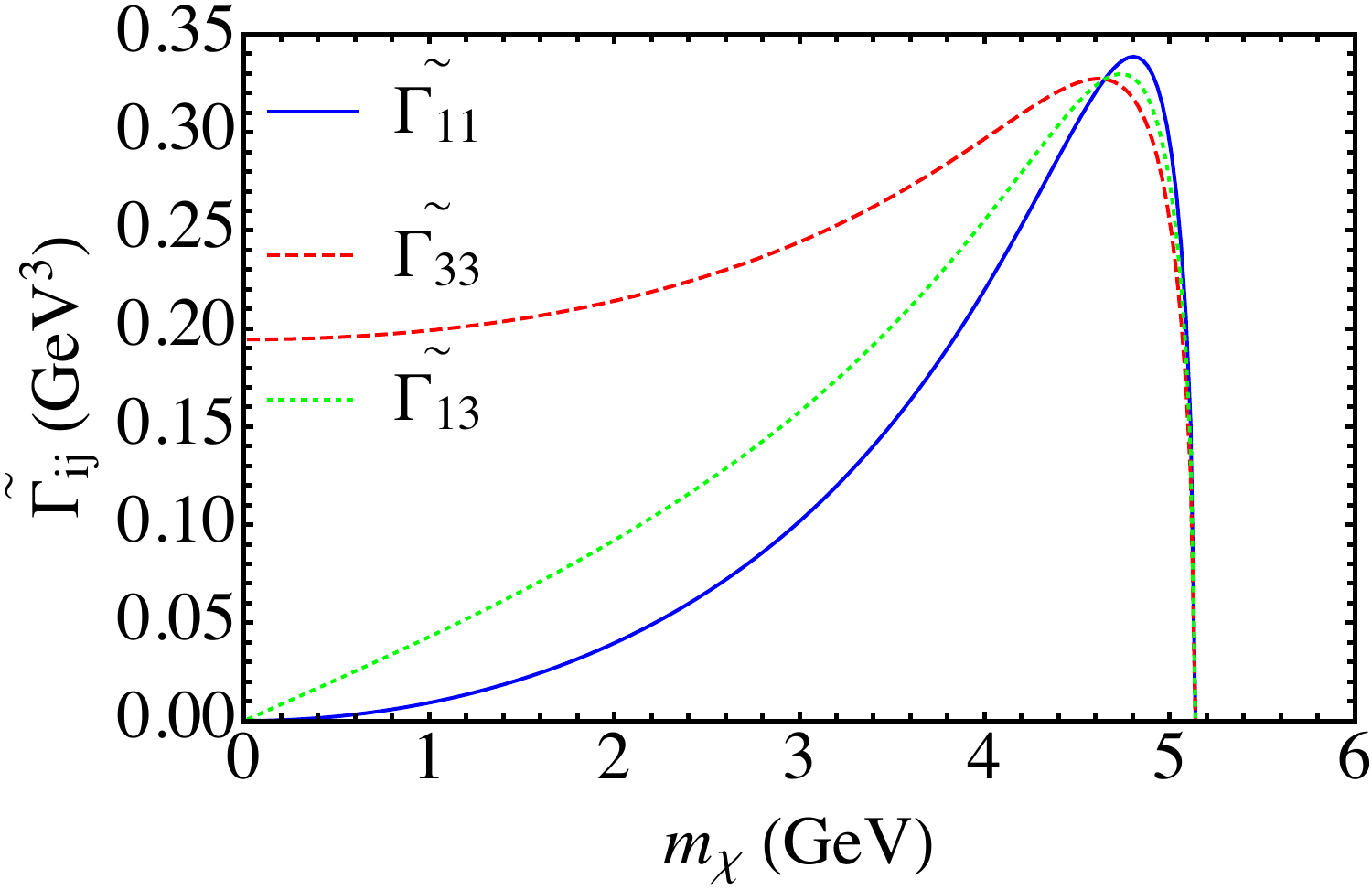}} \\
	\subfigure[$B_c^-\rightarrow D_s^-S$]{		\label{width-S14c}
		\includegraphics[width=0.45\textwidth]{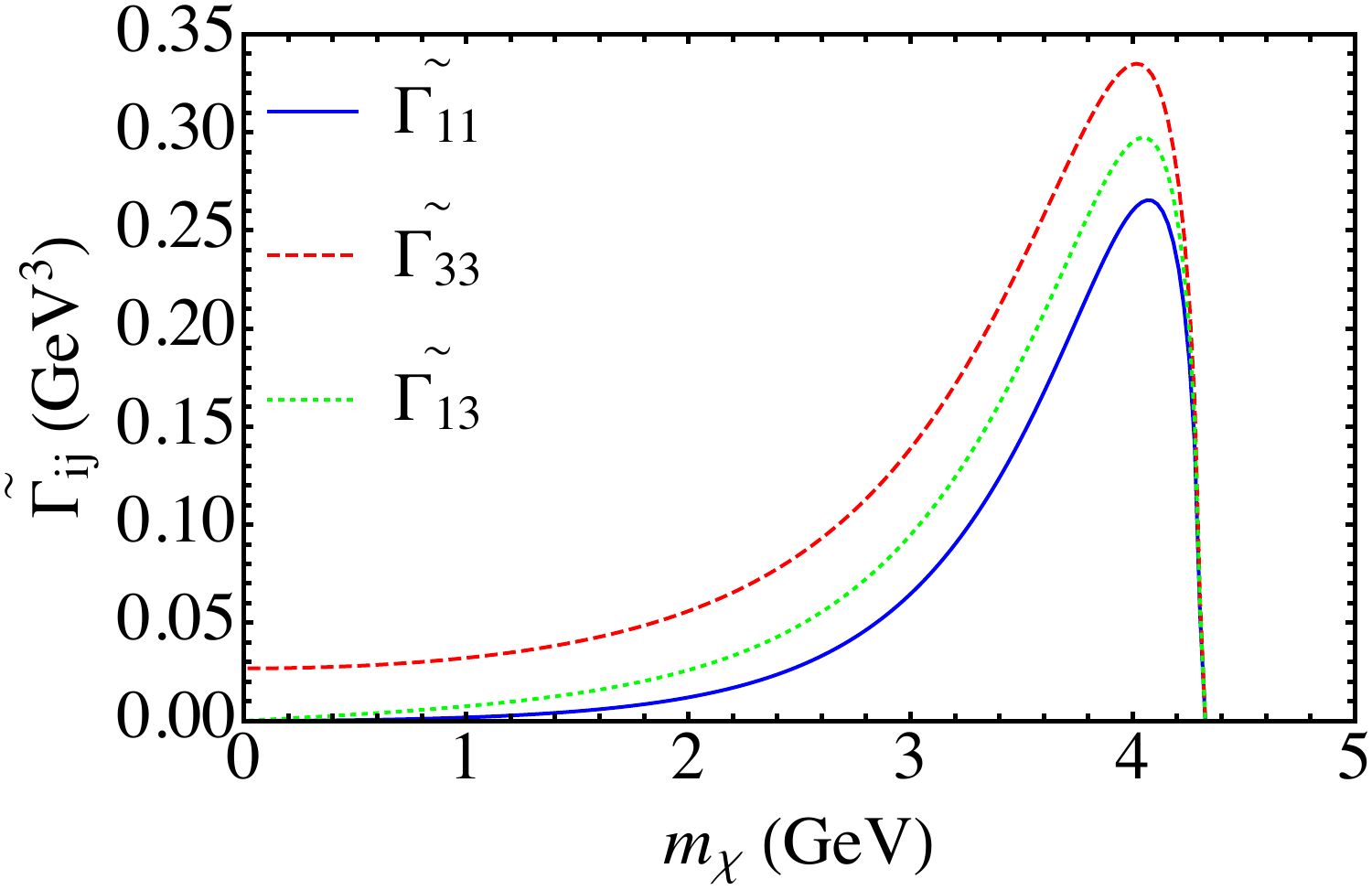}}
	\hspace{2em}
	\subfigure[$B_c^-\rightarrow D^-S$]{		\label{width-S14d}
		\includegraphics[width=0.45\textwidth]{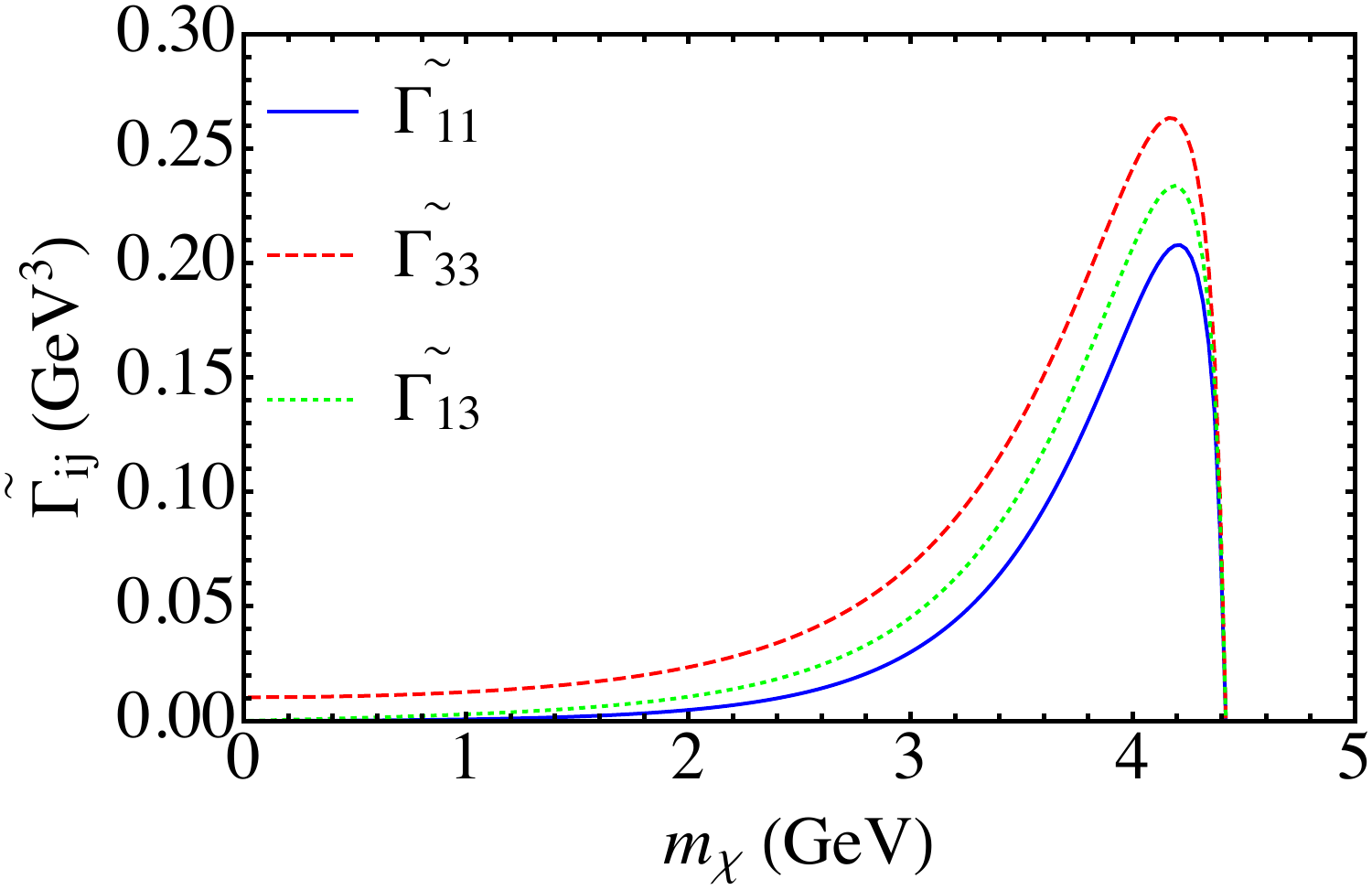}}
	\caption{$\widetilde\Gamma_{ij}$s in $B$ and $B_c$ meson $0^-\to 0^-$ decays with invisible scalar.}
	\label{width-S14}
\end{figure}

The results of $\widetilde\Gamma_{ij}$s are shown in Fig.~\ref{width-S14}. The solid and dashed lines represent noninterference and interference terms, respectively. 
One can see that although we use different parametric methods in $B$ and $B_c$ meson decays, the trends of $\widetilde\Gamma_{ij}$s are similar. This is because the mesons have same quantum numbers.  $\widetilde\Gamma_{ij}$ increases when $m_{_S}$ increases from zero. It grows faster in $B\to M_f$ modes than that in $B_c \to M_f$ modes when $m_{_S}$ is about smaller than $4$~GeV. This is due to the difference in the form factor caused by the masses of final state mesons, since $K$ and $\pi$ mesons are light while $D_{(s)}^*$ are heavy. $\widetilde\Gamma_{33}$ and $\widetilde\Gamma_{13}$ are zero when $m_{_S}\to0$, because they are proportional to $m_{_S}^2$ and $m_{_S}$, respectively. When $m_{_S}$ is about larger than $4$~GeV, $\widetilde\Gamma_{ij}$ decreases. When $m_{_S}=(M-M_f)$, $\widetilde\Gamma_{ij}$s are zero for there is no phase space. 

\begin{figure}[h]
	\centering
	\subfigure[$B^-\rightarrow K^-S$]{
		\includegraphics[width=0.45\textwidth]{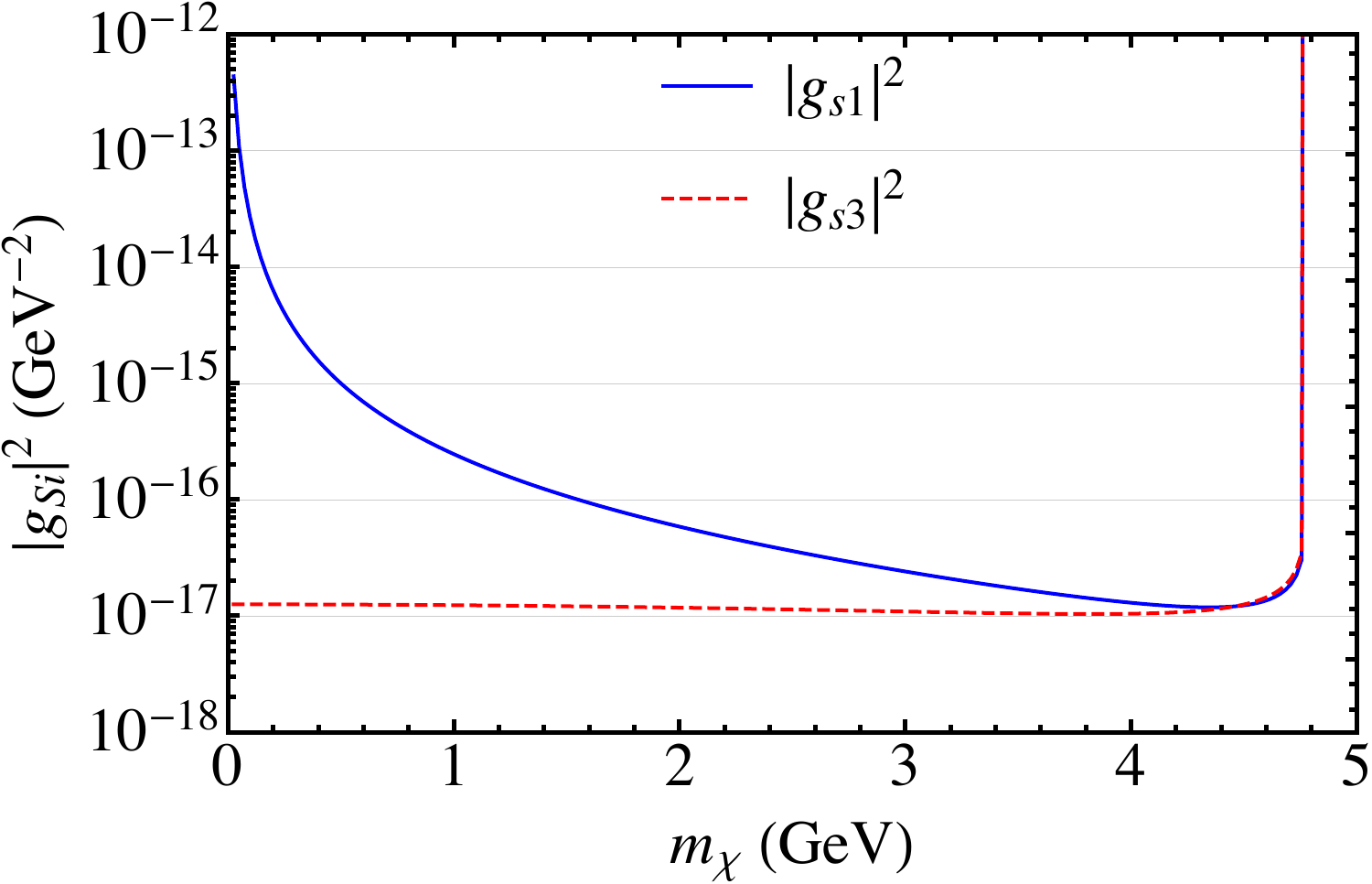}}
	\hspace{2em}
	\subfigure[$B^-\rightarrow \pi^-S$]{
		\includegraphics[width=0.45\textwidth]{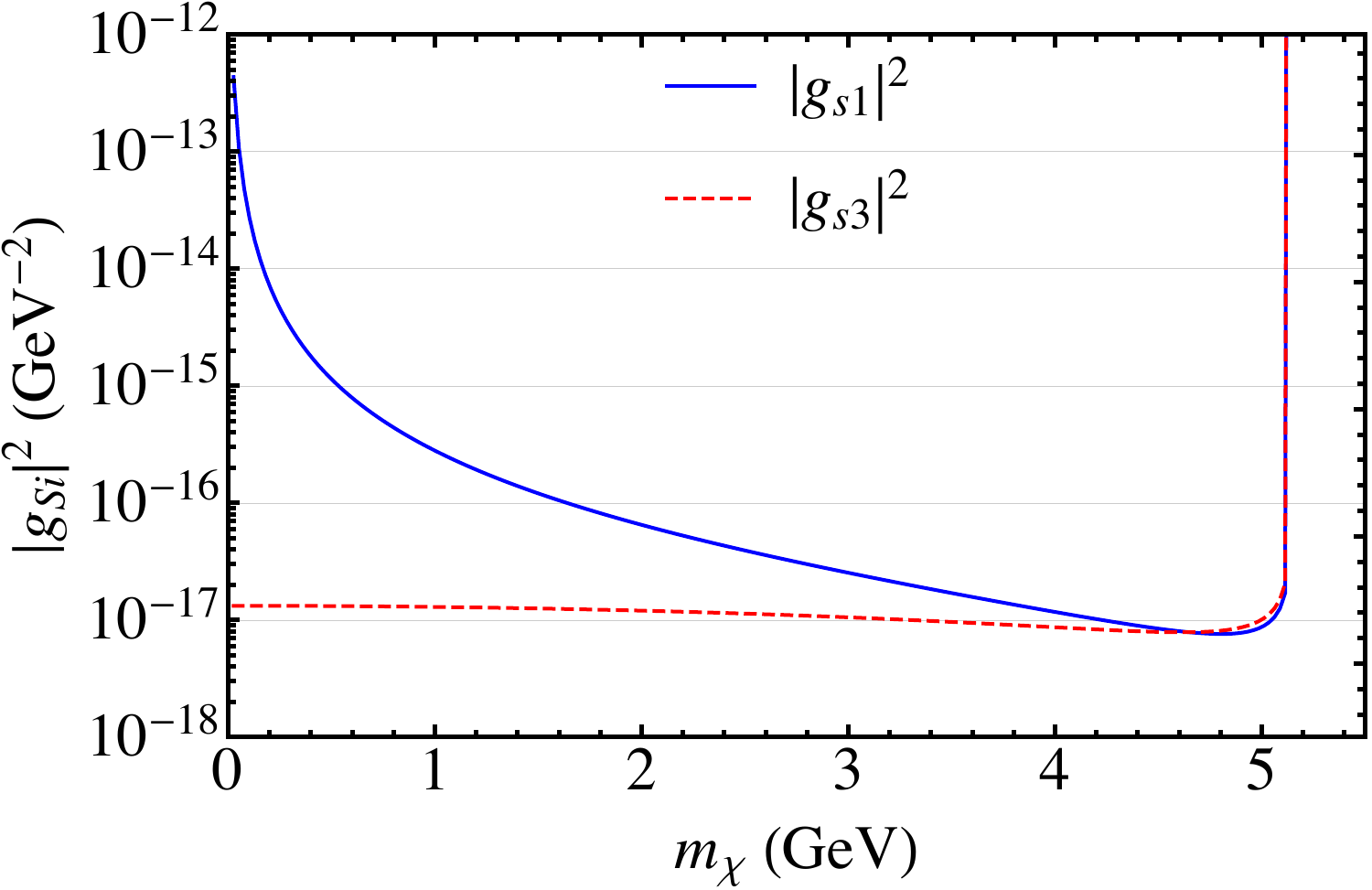}} 
	\caption{Upper limits of $g_{_{Si}}$s from $B$ meson $0^-\to 0^-$ decays with invisible scalar.}
	\label{gs13}
\end{figure}

\begin{figure}[h]
	\centering
	\subfigure[$B_c^-\rightarrow D_s^{-}S$]{
		\includegraphics[width=0.45\textwidth]{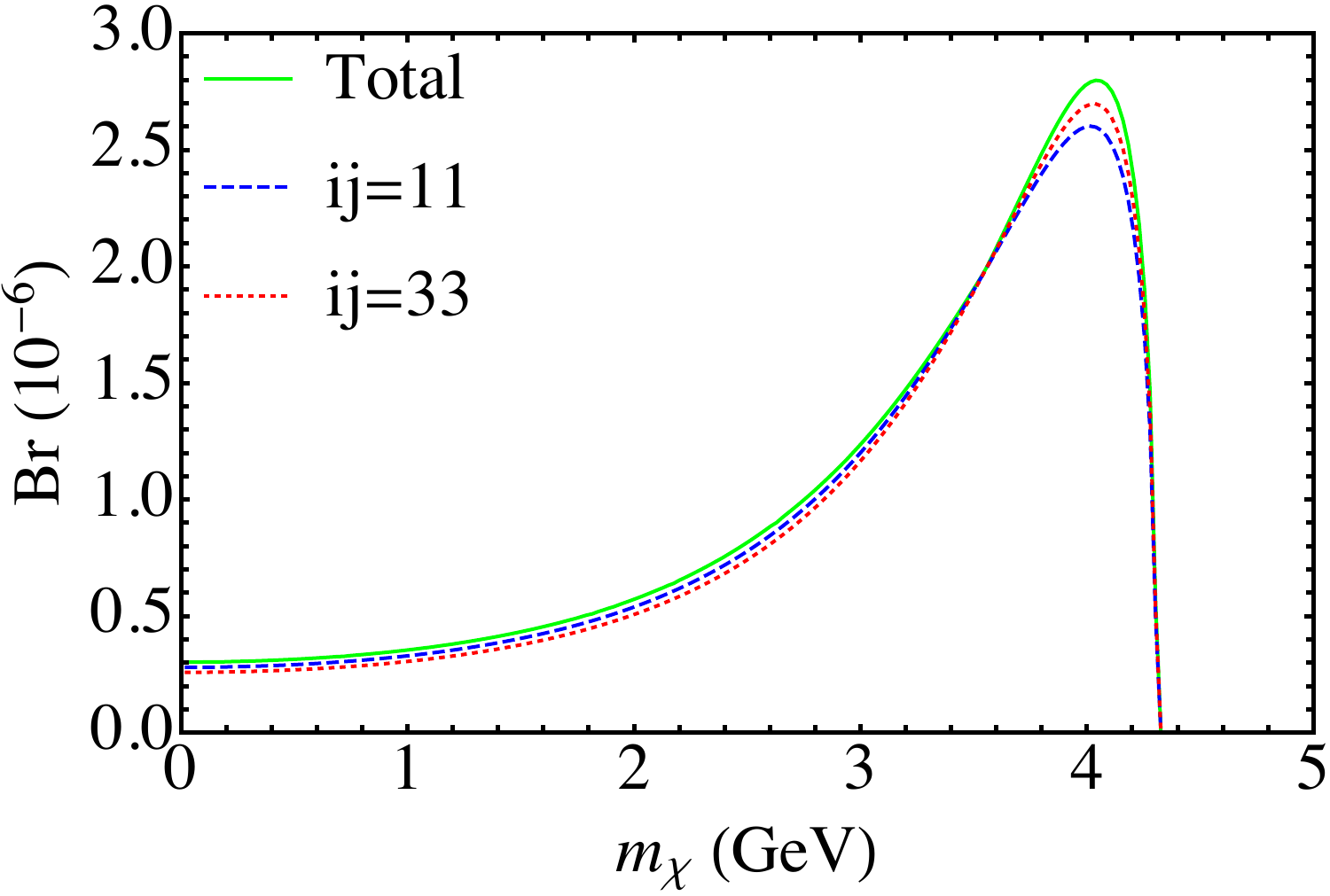}}
	\hspace{2em}
	\subfigure[$B_c^-\rightarrow D^{-}S$]{
		\includegraphics[width=0.45\textwidth]{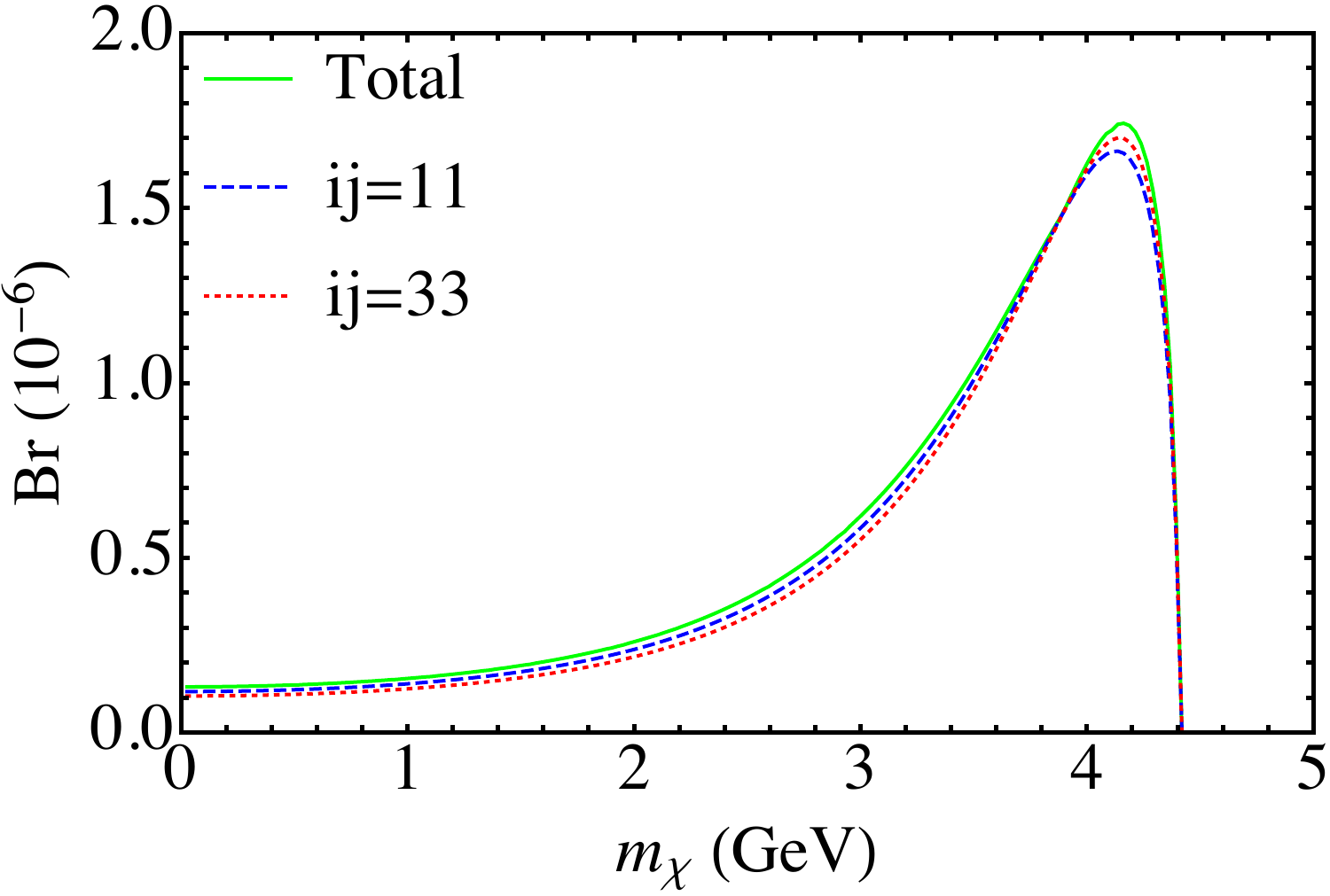}}
	\caption{Branching ratios of $B_c$ meson $0^-\to0^-$ decays with invisible scalar.}
	\label{br-S12}
\end{figure}

The upper limits in Table~\ref{tab1} give the allowed parameter space for the effective coupling constants $g_{_{Si}}$s. Here we use two different ways to make the calculation. First, we assume that only one of the $g_{_Si}$ is not zero and make others zero. In this case, the upper limits of $g_{_{Si}}$s as functions of $m_{_S}$ are shown in Fig.~\ref{gs13}, where the point of $m_{_S}=0$ is excluded. 
One can see that the upper limit of $|g_{_{Si}}|^2$ is infinite when $m_{_S}=M-M_f$. This is because $\widetilde\Gamma_{ij}=0$ at this point. The smallest valve of $|g_{_{Si}}|^2$ is of the order of $10^{-17}~{\rm GeV}^{-2}$. The solid blue line which represents $|g_{_{S1}}|^2$ is infinite when $m_{_S}\to 0$, since the blue solid line in Fig.~\ref{width-S14a} and Fig.~\ref{width-S14b} which represents $\widetilde\Gamma_{11}$ is zero at this point. The red dashed line which represents $|g_{_{S3}}|^2$ changes slowly when $m_{_S} \textless M-M_f$ due to $\widetilde\Gamma_{33}$ changes slowly in Fig.~\ref{width-S14a} and Fig.~\ref{width-S14b}. 
Second, we assume that all operators make contribution and run a program to select the maximum value of the branching ratio of $B_c$ meson. The results are plotted as dashed (ij=11, 33) and solid (Total) lines in Fig.~\ref{br-S12}, respectively. 
One can see that the upper limits of $\mathcal BR$ are of the order of $10^{-6}$. The results of two methods show subtle differences.

It should be noticed that these results are the upper limits of the branching ratios. The area under the curves in Fig.~\ref{br-S12} represents the possible values of the branching ratios. The peak is located near $m_{_S} \approx 4$~GeV, which may imply the greatest probability of detecting the invisible particles in this area. Taking the LHC as an example, although the generation of $B_c$ meson cases can reach the order of $10^{10}$ per year, the actual effective detection is still several orders of magnitude lesser. If more cases can be detected and the distribution spectrum of the missing energy can be obtained, then it is possible to observe the signal experimentally. 

\subsection{$0^-\to 1^-$ meson decay processes}

\begin{figure}[h]
	\centering
	\subfigure[$B^-\rightarrow K^{*-}S$]{
		\includegraphics[width=0.45\textwidth]{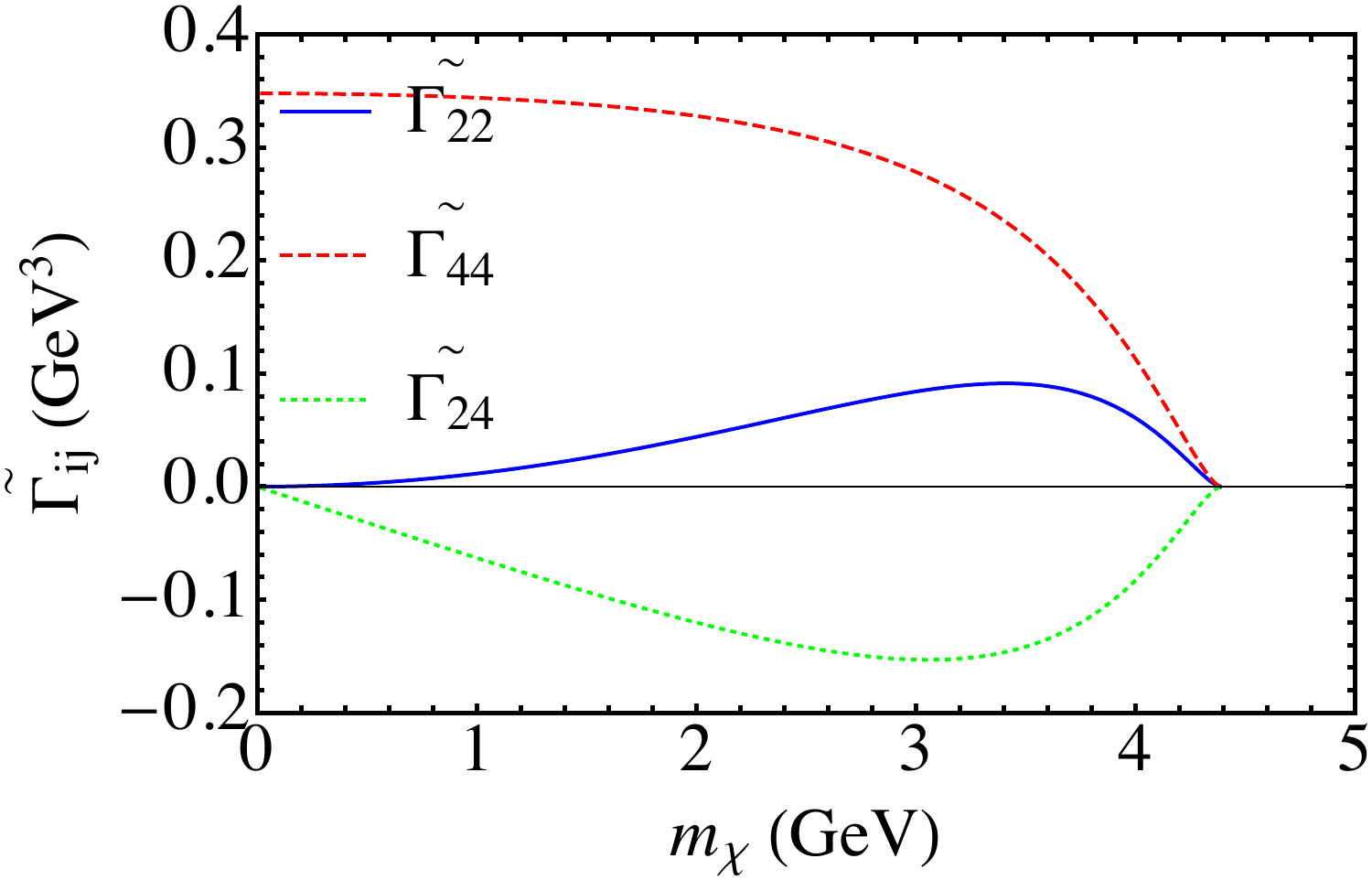}}
	\hspace{2em}
	\subfigure[$B^-\rightarrow \rho^-S$]{
		\includegraphics[width=0.45\textwidth]{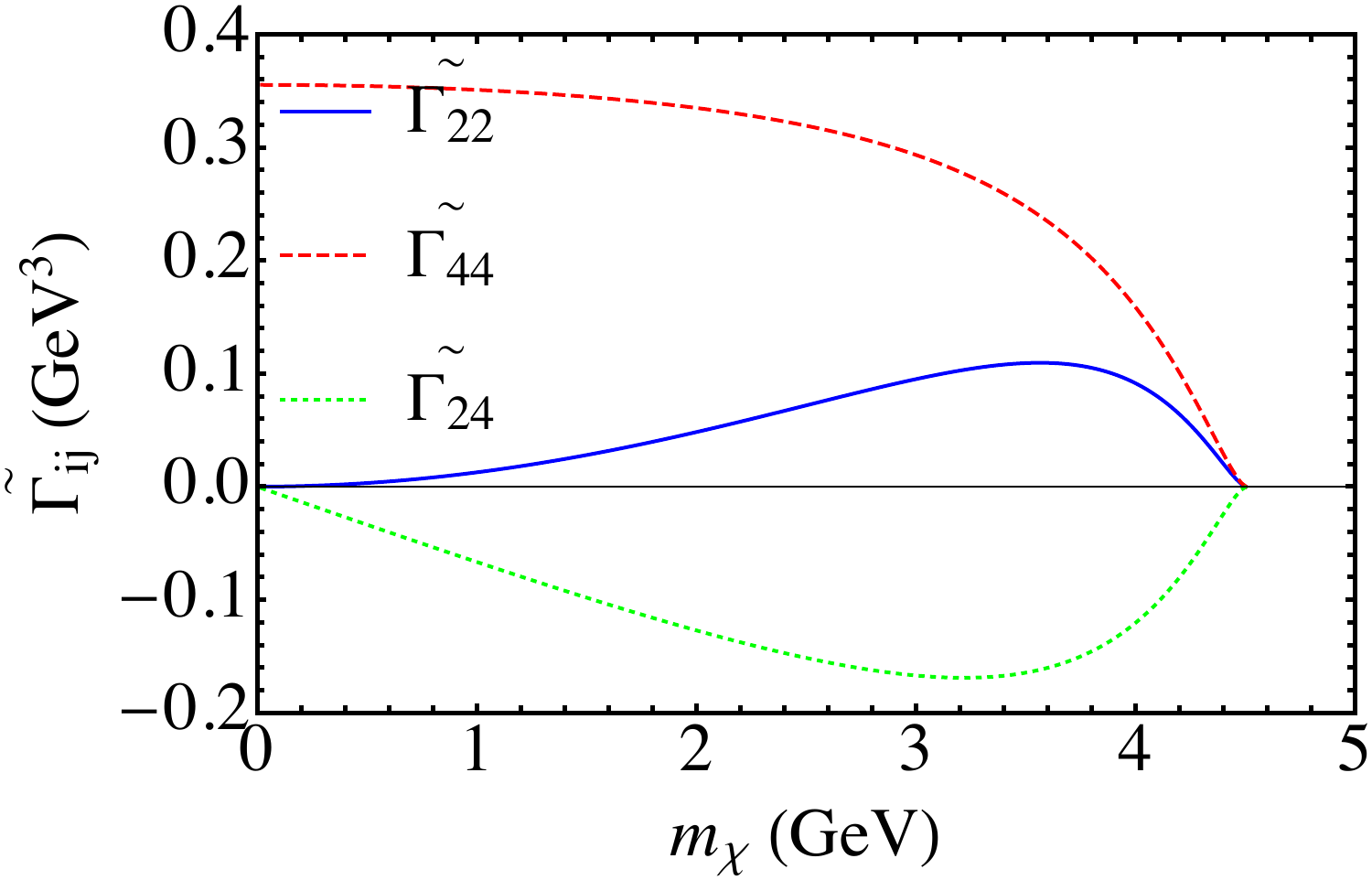}} \\
	\subfigure[$B_c^-\rightarrow D_s^{*-}S$]{
		\includegraphics[width=0.45\textwidth]{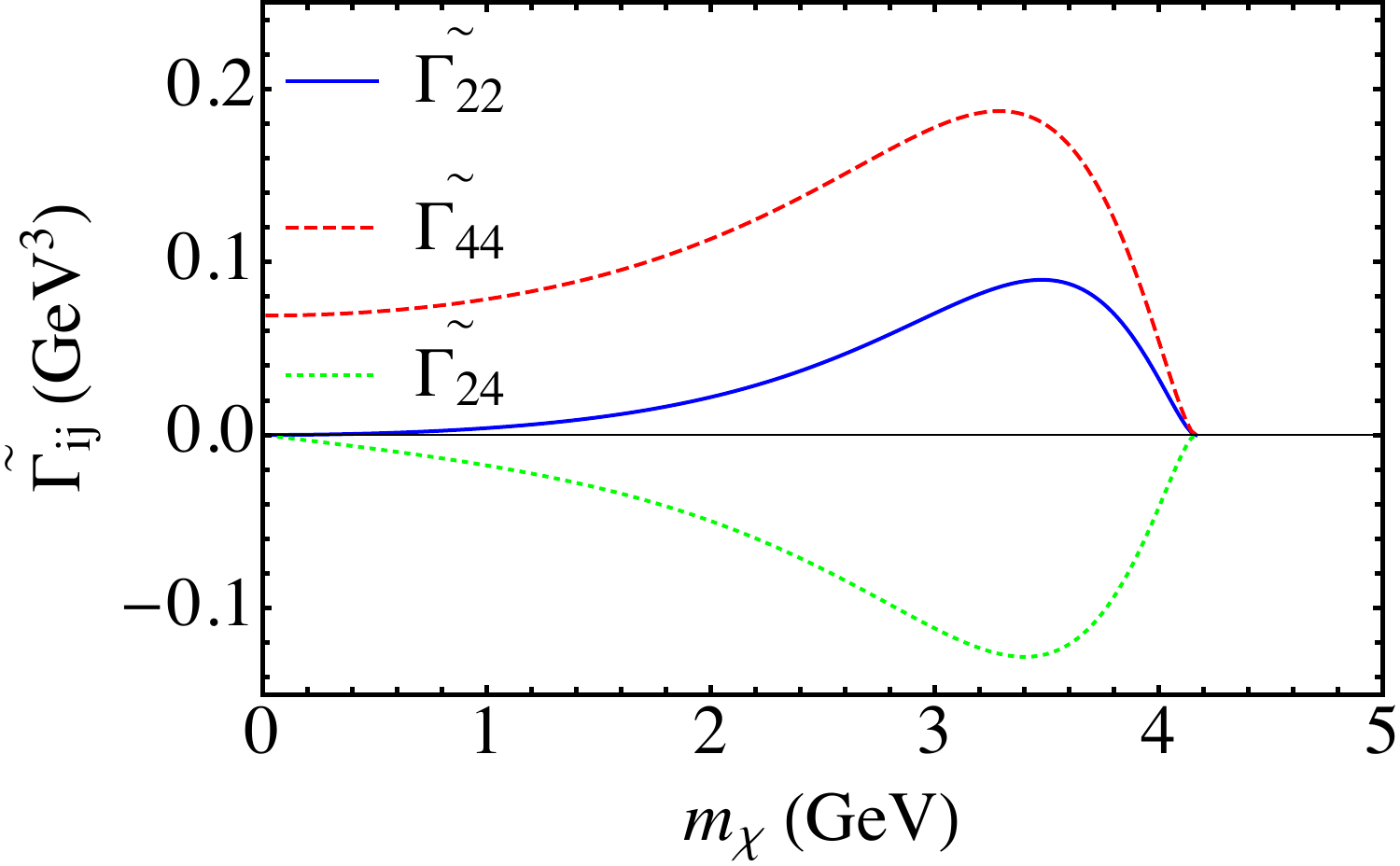}}
	\hspace{2em}
	\subfigure[$B_c^-\rightarrow D^{*-}S$]{
		\includegraphics[width=0.45\textwidth]{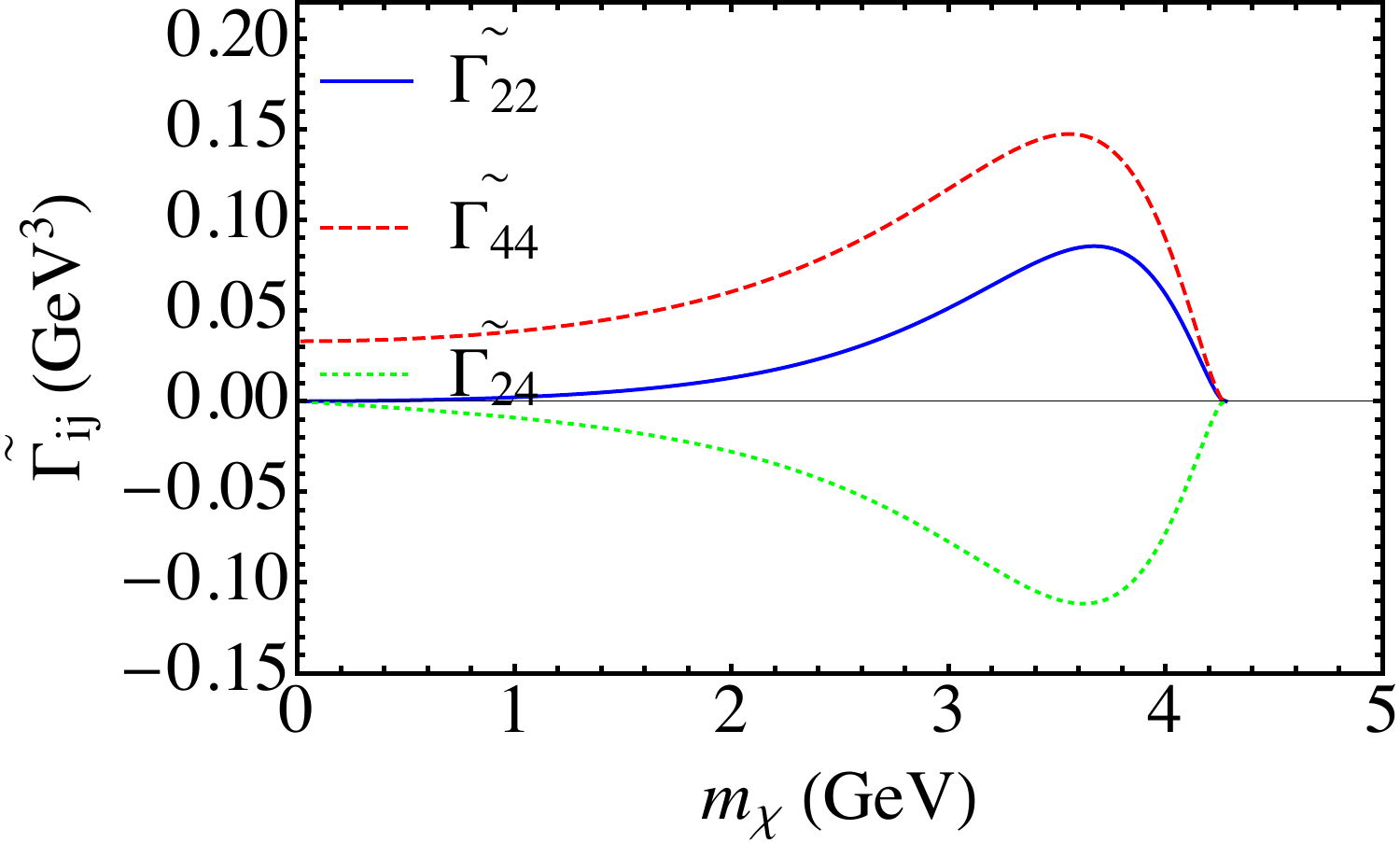}}
	\caption{$\widetilde\Gamma_{ij}$s in $B$ and $B_c$ meson $0^-\to 1^-$ decays with invisible scalar.}
	\label{width-S58}
\end{figure}

In $0^-\to1^-$ meson decays, the decay width has the form
\begin{equation}
	\begin{aligned}
		\Gamma(M\to M^{*}_f S)=& \frac{1}{16\pi M^3 } \lambda^{1/2}(M^2,M_f^{*2},m_{_S}^2) \bigg\{ m_{_S}^2 g_{_{S2}}^2\langle M_f^{*-}|(\bar q_{_f} \gamma^5 q)|M^-\rangle^*\langle M_f^{*-}|(\bar q_{_f} \gamma^5 q)|M^-\rangle \\
		&+g_{_{S4}}^2 \langle M_f^{*-}|(\bar q_{_f} \gamma_{\nu} \gamma^5 q)|M^-\rangle^*\langle M_f^{*-}|(\bar q_{_f}\gamma_{\mu} \gamma^5 q)|M^-\rangle P_{_S}^\nu P_{_S}^\mu \\
		&+m_{_S} g_{_{S2}}g_{_{S4}}^*\langle M_f^{*-}|(\bar q_{_f} \gamma_{\nu} \gamma^5 q)|M^-\rangle^* \langle M_f^{*-}|(\bar q_{_f} \gamma^5 q)|M^-\rangle P_{_S}^\nu \bigg\},
		\label{eq6}
	\end{aligned}
\end{equation}
where $M_f^*$ represents the mass of $1^-$ final meson. The hadronic transition matrix elements can be expressed as the functions of form factors
\begin{equation}
	\begin{aligned}
		\langle M_f^{*-}|(\bar q_f \gamma^5 q )|M^-\rangle
		&\simeq -\frac{(P-P_f)^\mu}{m_q+m_{q_{_f}}}\langle M_f^{*-}|(\bar q_{_f}\gamma_\mu\gamma^5 q) |M^-\rangle =-i\frac{2M_f^*}{m_q+m_{q_{_f}}}\epsilon \cdot (P-P_f)A_0(s)\\
		\langle M_f^{*-}|(\bar q_f \gamma_\mu q )|M^-\rangle
		&=\varepsilon _{\mu \nu \rho \sigma} \epsilon ^\nu P^\rho (P-P_f)^\sigma\frac{2 }{M+M_f^*}V(s), \\
		\langle M_f^{*-}|(\bar q_{_f}\gamma_\mu\gamma^5 q) |M^-\rangle
		&= i\bigg\{\epsilon_{\mu} (M+M_f^*)A_1(s)-(P+P_f)_{\mu}\frac{\epsilon \cdot (P-P_f)}{M+M_f^*}A_2(s)\\
		&~~~-(P-P_f)_{\mu}  \big[\epsilon \cdot (P-P_f)\big] \frac{2M_f^*}{s} \big[A_3(s)-A_0(s)\big]\bigg\}.
		\label{eq7}
	\end{aligned}
\end{equation}
where the parameters are cited from LCSR method~\cite{Straub:2015ica} in $B$ meson decays. The BS method~\cite{Kim:2003ny, Wang:2005qx} is applied to calculate the hadronic transition matrix element of $B_c$ meson decays. The results of $\widetilde\Gamma_{ij}$s are shown in Fig.~\ref{width-S58}.
It can be seen that there is an obvious difference between $B$ and $B_c$ meson. As $m_{_S}$ increases, $\widetilde\Gamma_{44}$ in $B$ meson decay increases first until $m_{_S}\approx 3.5$~GeV, then decreases to zero. While $\widetilde\Gamma_{44}$ in $B_c$ meson decay keep deceasing until there is no phase space. This is the result of competition between form factors and phase space. As $m_{_S}$ increases, the form factors increase while phase space decreases. The form factors of $B_c$ meson decays grow much faster than these of the $B$ meson decays.

\begin{figure}[h]
	\centering
	\subfigure[$B^-\rightarrow K^{*-}S$]{
		\includegraphics[width=0.45\textwidth]{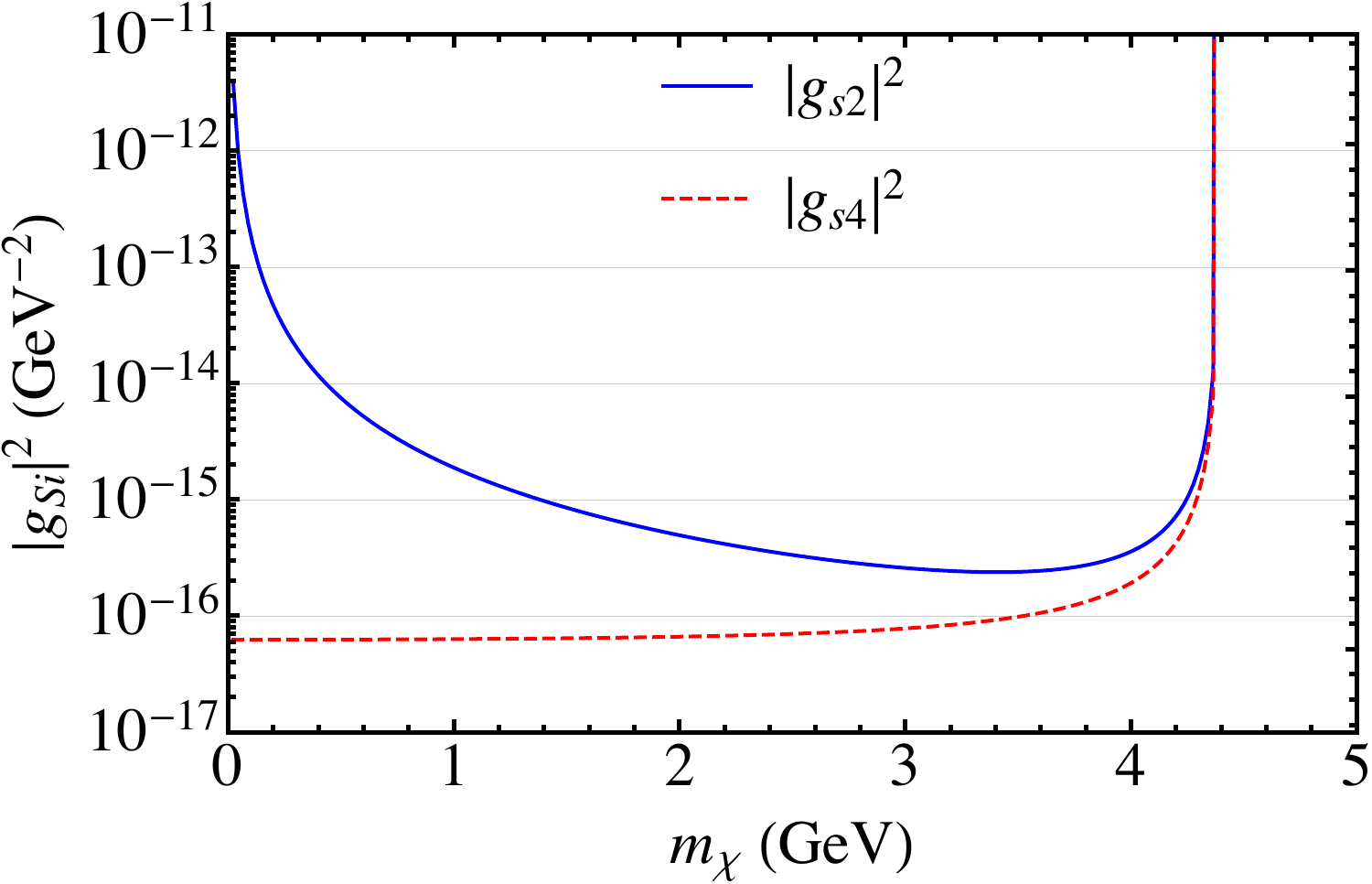}}
	\hspace{2em}
	\subfigure[$B^-\rightarrow \rho^-S$]{
		\includegraphics[width=0.45\textwidth]{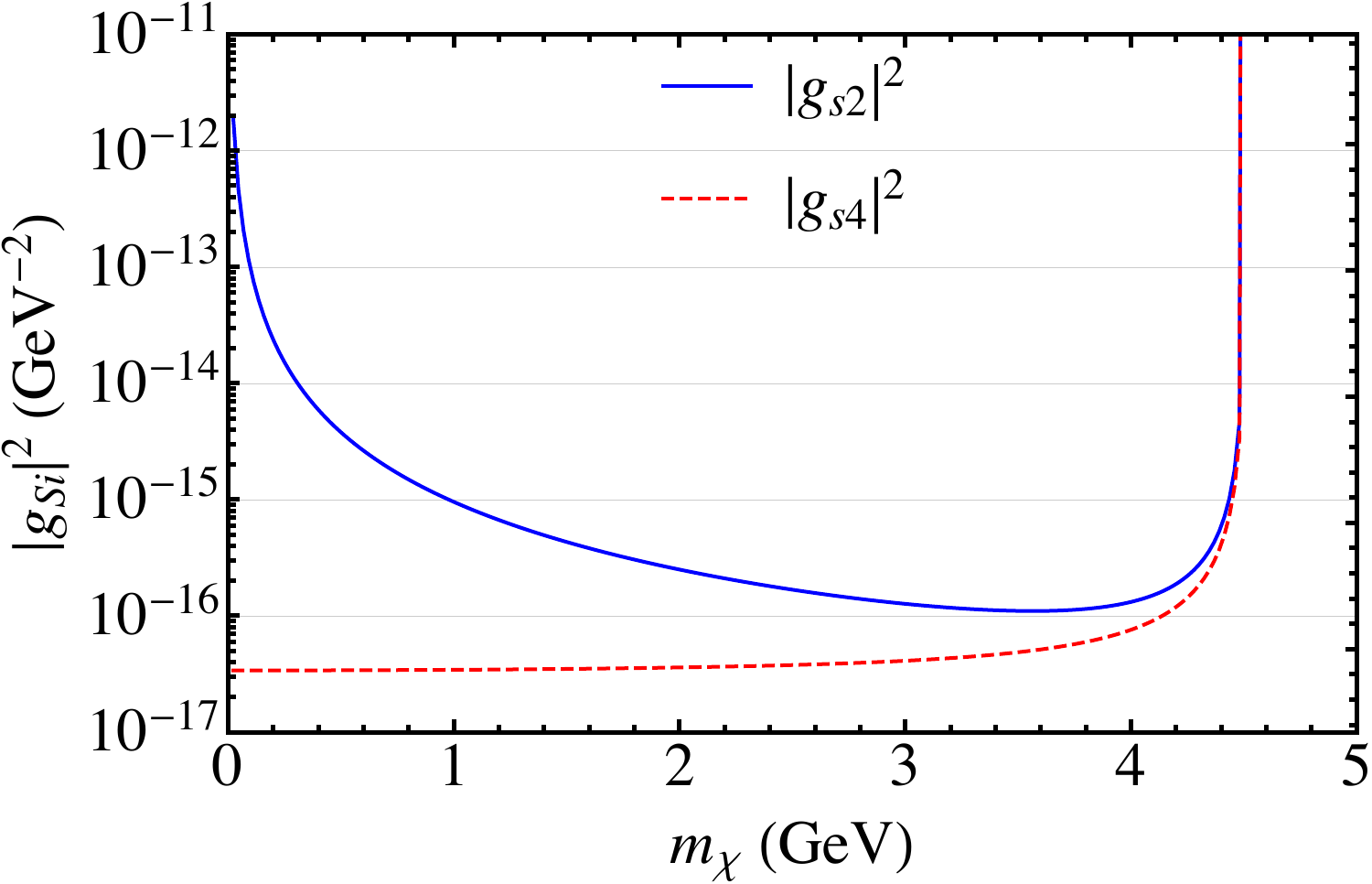}} 
	\caption{Upper limits of $g_{_{Si}}$s from $B$ meson $0^-\to 1^-$ decays with invisible scalar.}
	\label{gs24}
\end{figure}

We also use two ways to set the upper limits for the branching ratios of $B_c^-\to M_f^{*-}S$ processes. The upper limits of $g_{_{Si}}$s obtained by the first method are shown in Fig.~\ref{gs24}. 
One can see that they have very similar trends to those in Fig.~\ref{gs13}, but about one order of magnitude bigger. This is caused by the different upper limits of experiments in Table~\ref{tab1}. 

\begin{figure}[h]
	\centering
	\subfigure[$B_c^-\rightarrow D_s^{*-}S$]{
		\includegraphics[width=0.45\textwidth]{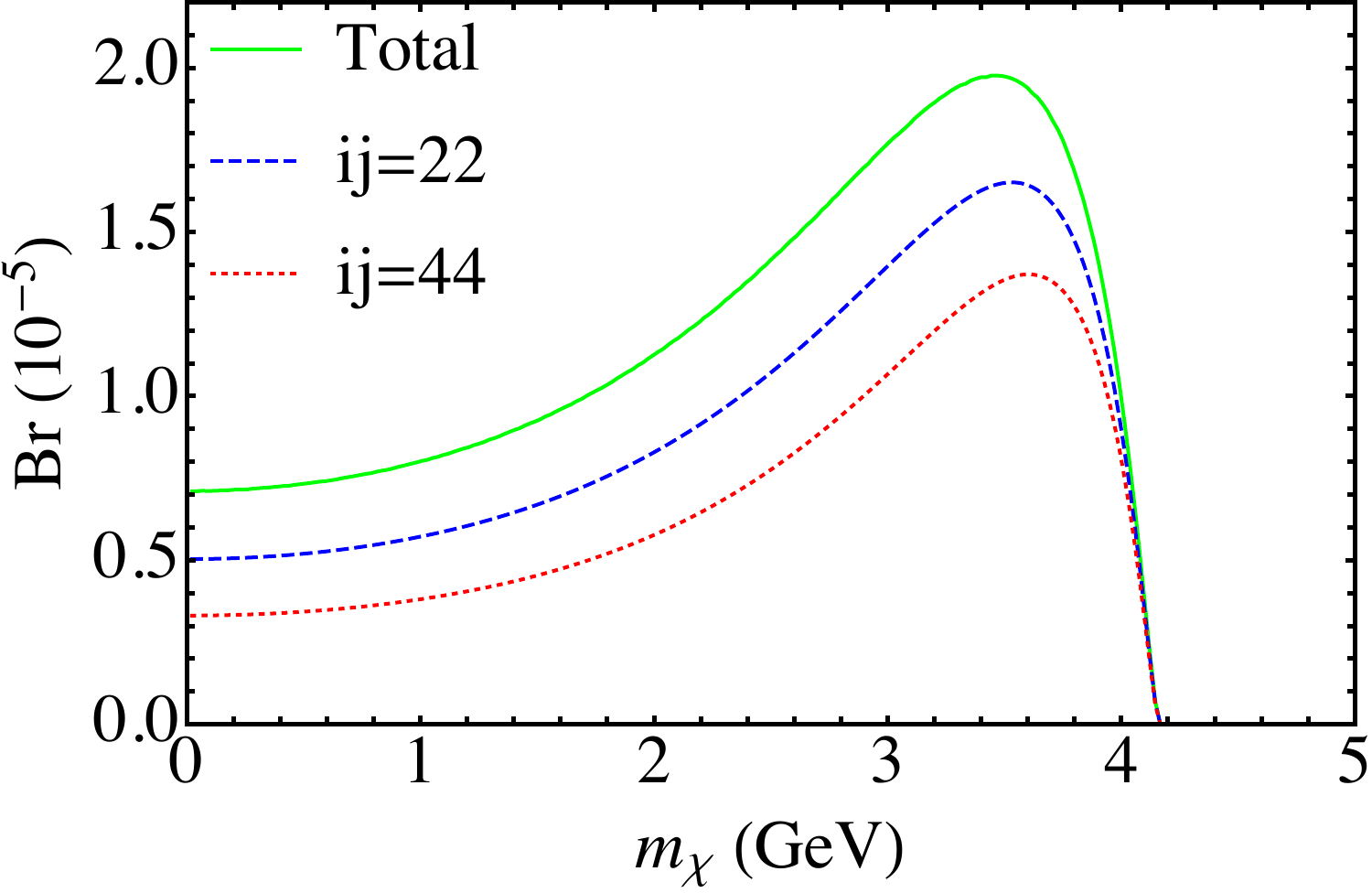}}
	\hspace{2em}
	\subfigure[$B_c^-\rightarrow D^{*-}S$]{
		\includegraphics[width=0.45\textwidth]{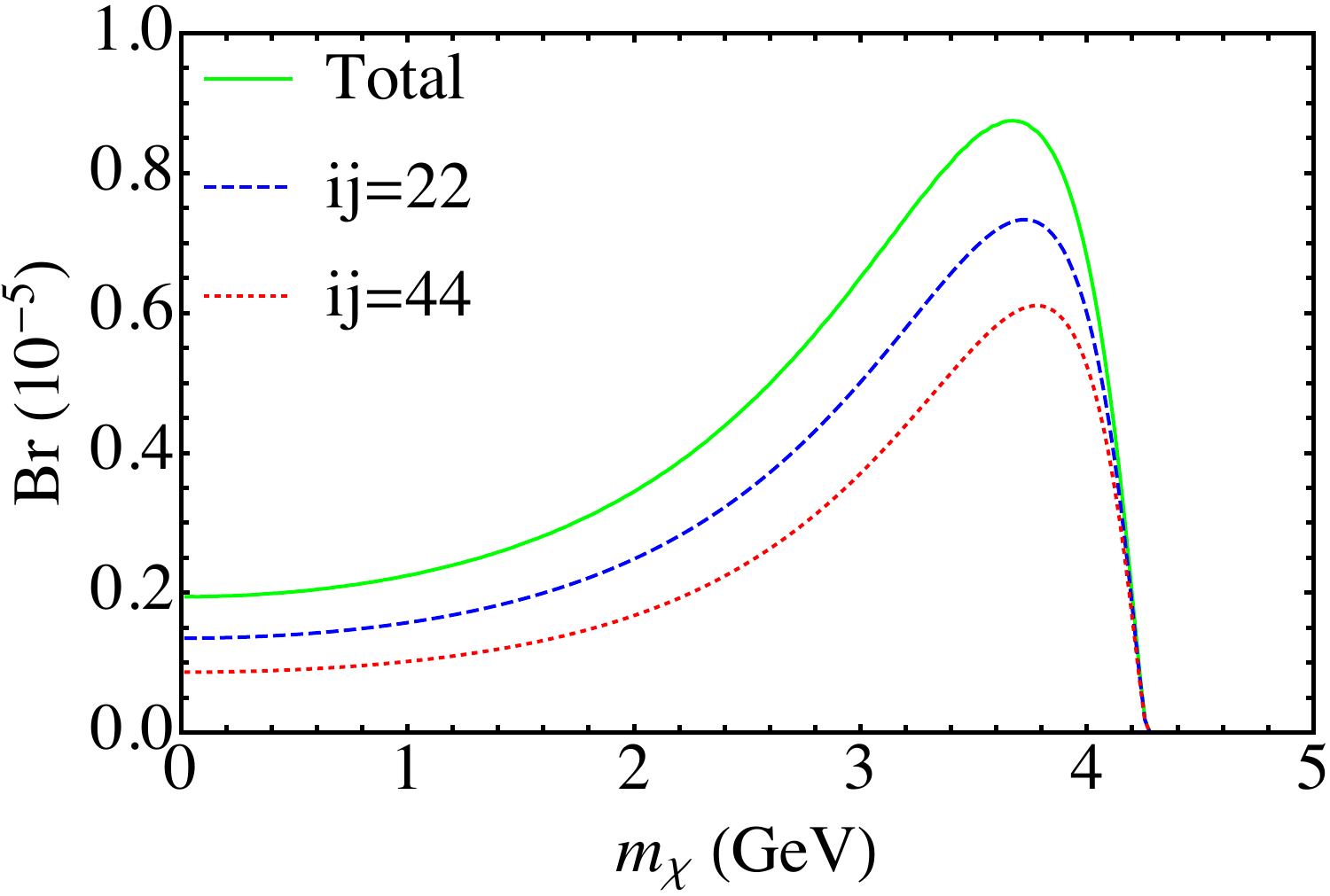}}
	\caption{Branching ratios of $B_c$ meson $0^-\to1^-$ decays with invisible scalar.}
	\label{br-S34}
\end{figure}

The upper limits of branching ratios of $B_c$ meson from two methods are shown in Fig.~\ref{br-S34}. 
One can see that the difference between dashed and solid lines are obvious. It is due to the contribution of interference term $\Gamma_{24}$. The most likely area for finding the dark scalar is near $m_{_S}\approx 3.5$~GeV. The $\mathcal {BR}$ is of the order of $10^{-5}$, which is about an order of magnitude larger than that in $0^-\to0^-$ modes. This depends on the experimental upper limits in Table~\ref{tab1}.

\section{Light invisible vector}

When $\chi=V$, we assume a hidden vector produced in the FCNC processes. The effective Lagrangian, which represents the coupling between SM fermions and the hidden vector, has the form
\begin{equation}
\begin{aligned}
\mathcal L_{vector}=m_{_V} g_{_{V1}}(\bar q_{_f}\gamma_{\mu}q)V^\mu + m_{_V} g_{_{V2}}(\bar q_{_f} \gamma_{\mu}\gamma^{5}q)V^\mu.
\label{eq8}
\end{aligned}
\end{equation}
This dimension-5 effective Lagrangian naturally meets gauge symmetry, since the chirality of two quarks are the same.

\subsection{$0^-\to 0^-$ meson decay processes}

\begin{figure}[h]
	\centering
	\subfigure[$B^-\rightarrow K^-(\pi^-)V$]{
	\includegraphics[width=0.45\textwidth]{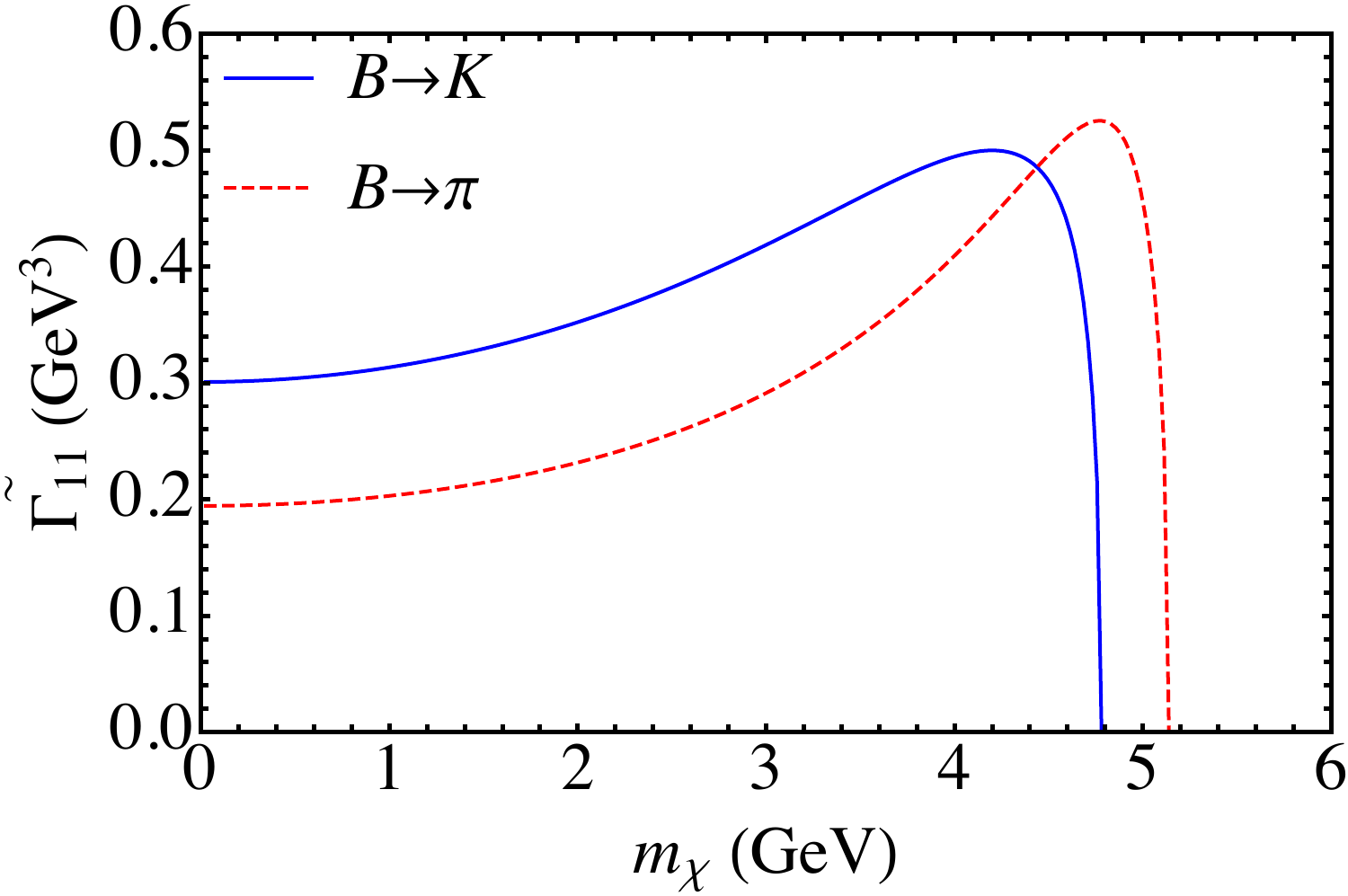}}
	\hspace{2em}
	\subfigure[$B_c^-\rightarrow D_{(s)}^-V$]{
	\includegraphics[width=0.45\textwidth]{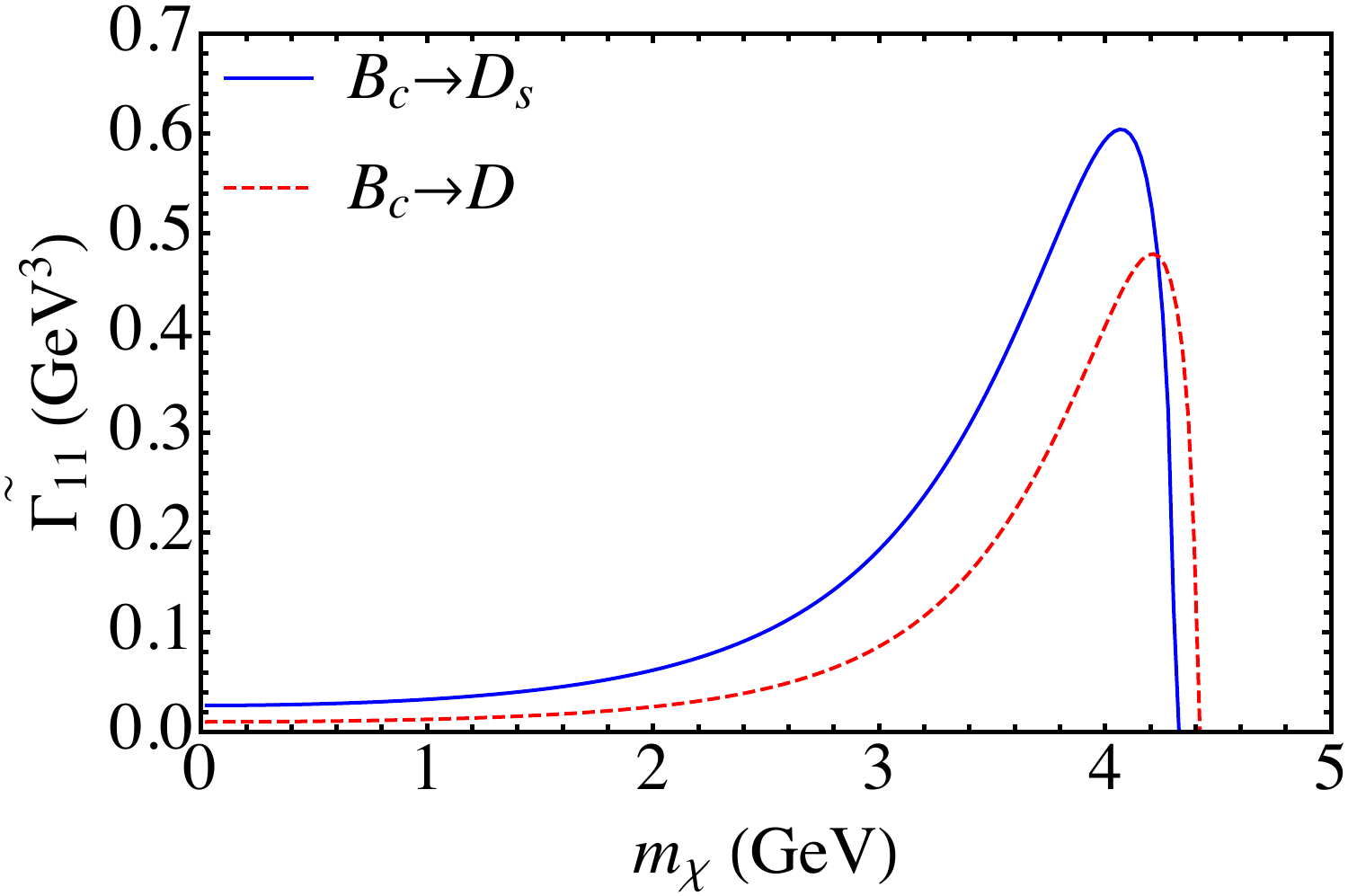}}
	\caption{$\widetilde\Gamma_{11}$ in $0^-\to 0^-$ meson decays with invisible vector.}
	\label{width-V12}
\end{figure}

By finishing the two-body phase space integral, the decay width of $M^- \to M_f^- V$ processes can be written as
\begin{equation}
	\begin{aligned}
		\Gamma(M\to M_f V )=\frac{m_{_V}^2 g_{_{V1}}^2}{16\pi M^3 } \lambda^{1/2}(M^2,M_f^2,m_{_V}^2)  \langle M_f^-|(\bar q_{_f} \gamma_{\nu}q)|M^-\rangle^*\langle M_f^-|(\bar q_{_f} \gamma_{\mu}q)|M^-\rangle \mathcal P_{_V}^{\mu\nu},
		\label{eq9}
	\end{aligned}
\end{equation}
where the sum of polarization vector is 
\begin{equation}
	\begin{aligned}
		\mathcal P_{_V}^{\mu\nu}=\sum\epsilon_{_V}^{*\mu}\epsilon_{_V}^{\nu}=-g^{\mu\nu}+\frac{P_{_V}^\mu P_{_V}^\nu}{m_{_V}^2}.
		\label{eq10}
	\end{aligned}
\end{equation}
The hadronic transition matrix element with pseudoscalar current is zero when final meson is pesudoscalar. The only nonzero term $\widetilde\Gamma_{11}$ is shown in Fig.~\ref{width-V12}. One can see that the results are smooth and convergent when $m_{_V}\to 0$. 

\begin{figure}[h]
	\centering
	\subfigure[Upper limits of $g_{_{V1}}$ from $B$ meson $0^-\to 0^-$ decays with invisible vector.]{
		\label{gv1}
		\includegraphics[width=0.46\textwidth]{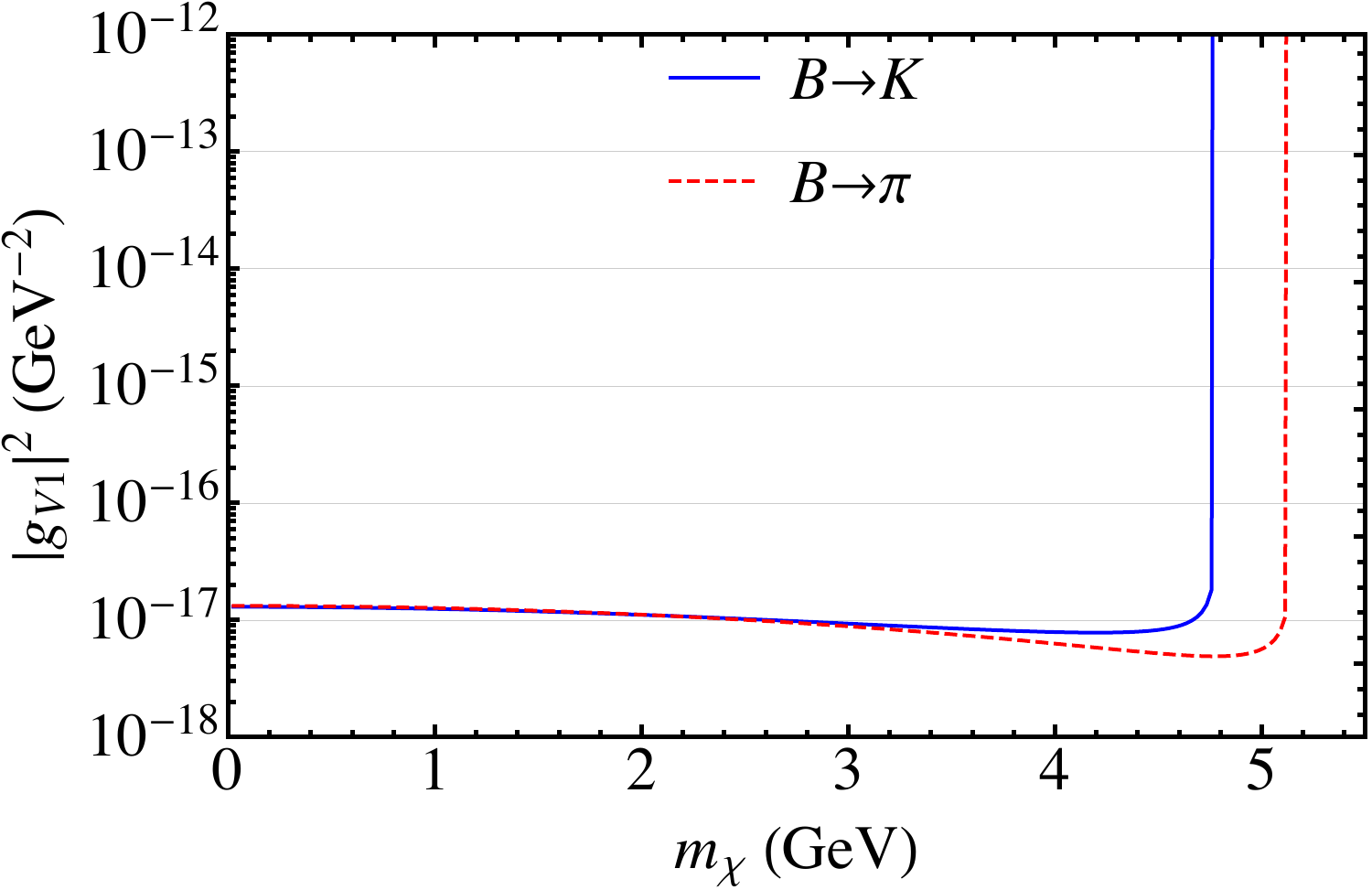}}
	\hspace{2em}
	\subfigure[Branching ratios of $B_c$ meson $0^-\to0^-$ decays with invisible vector.]{
		\label{br-V1}
		\includegraphics[width=0.43\textwidth]{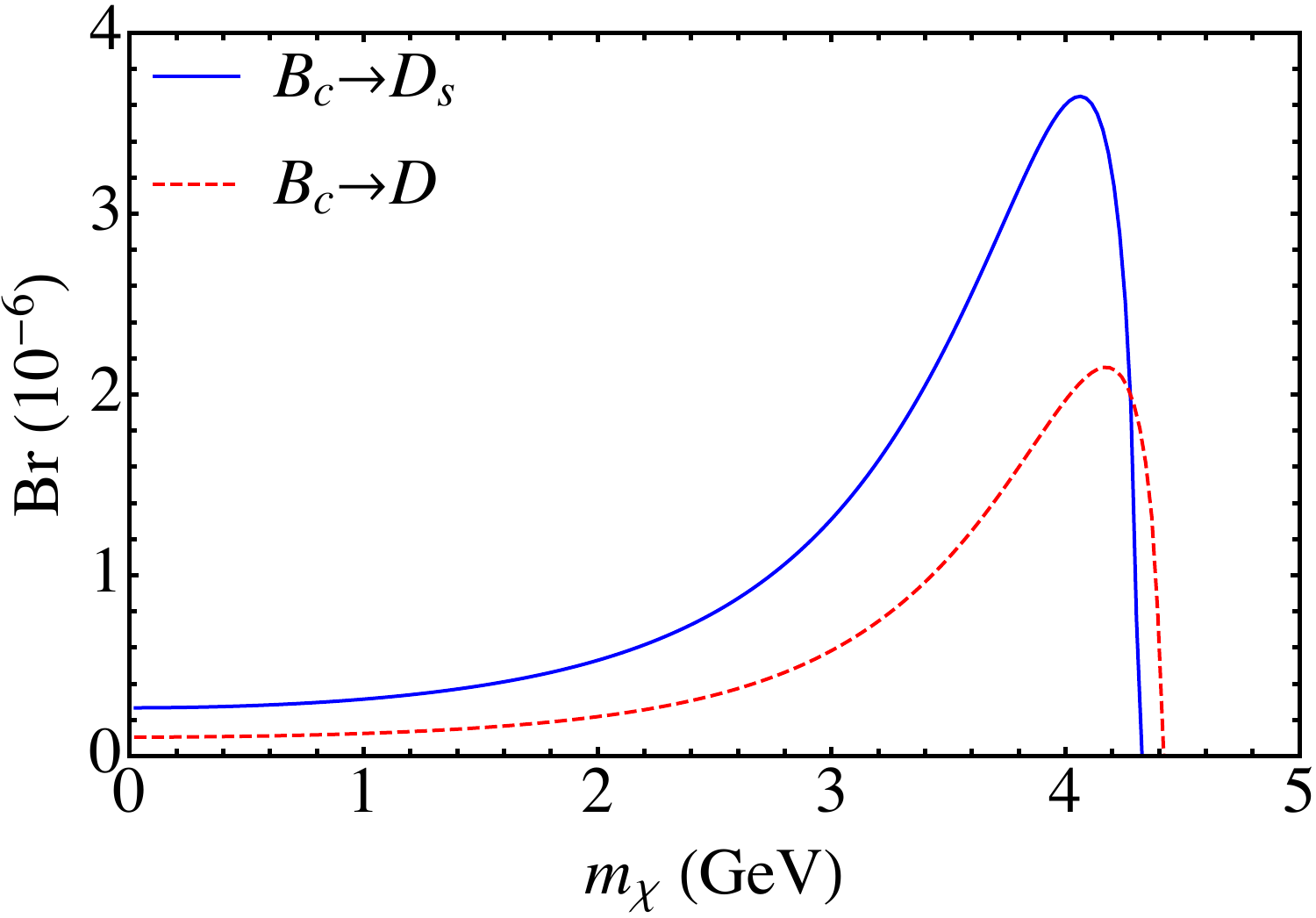}}
	\caption{Upper limits of coupling constants and $\mathcal {BR}$ of $B_c$ meson with invisible vector.}
	\label{upper}
\end{figure}

Since only one operator contributes, the upper limits of the coupling constants and branching ratios of $B_c$ meson can be easily obtained, which are shown in Fig.~\ref{upper}. 
One can see that the upper limits of $|g_{_{V1}}|^2$ are infinite when $m_{_V}=M-M_f$, since $\widetilde\Gamma_{11}=0$ at this point. It changes slowly when $m_{_V} \textless M-M_f$ because $\widetilde\Gamma_{11}$ changes slowly in Fig.~\ref{width-V12}. The upper limits of branching ratios are of the order of $10^{-6}$. As the mass of the invisible particle increases, the upper limits of $\mathcal {BR}$ increase first and then decrease to zero. The peak is located near $m_{_V}\approx 4$~GeV. It may be the area where the invisible particle is most likely to be detected experimentally.

\subsection{$0^-\to 1^-$ meson decay processes}

\begin{figure}[h]
	\centering
	\subfigure[$B^-\rightarrow K^{*-}(\rho^-)V$]{ \label{width-V34a}
		\includegraphics[width=0.45\textwidth]{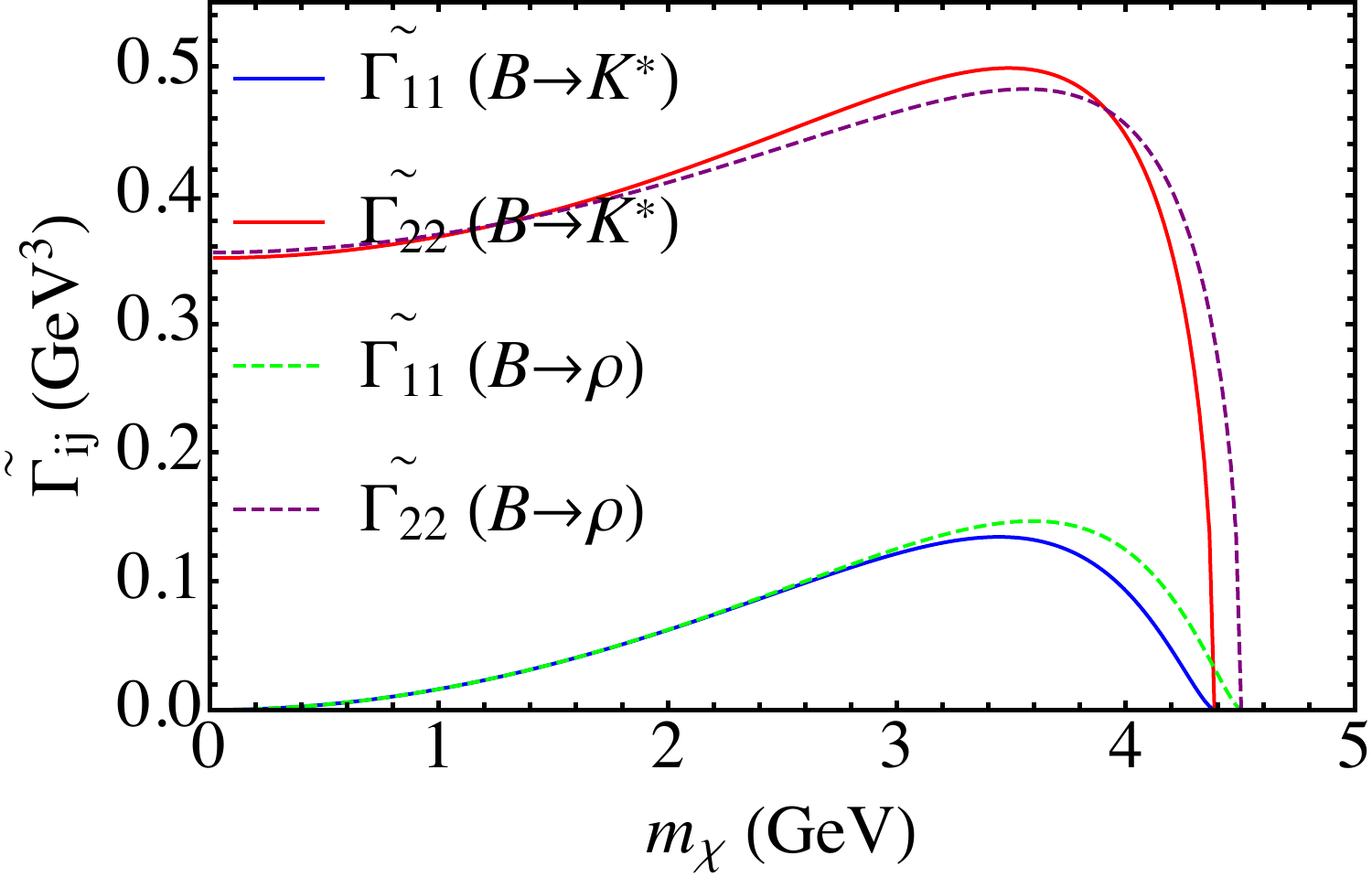}}
	\hspace{2em}
	\subfigure[$B_c^-\rightarrow D_{(s)}^{*-}V$]{ \label{width-V34b}
		\includegraphics[width=0.45\textwidth]{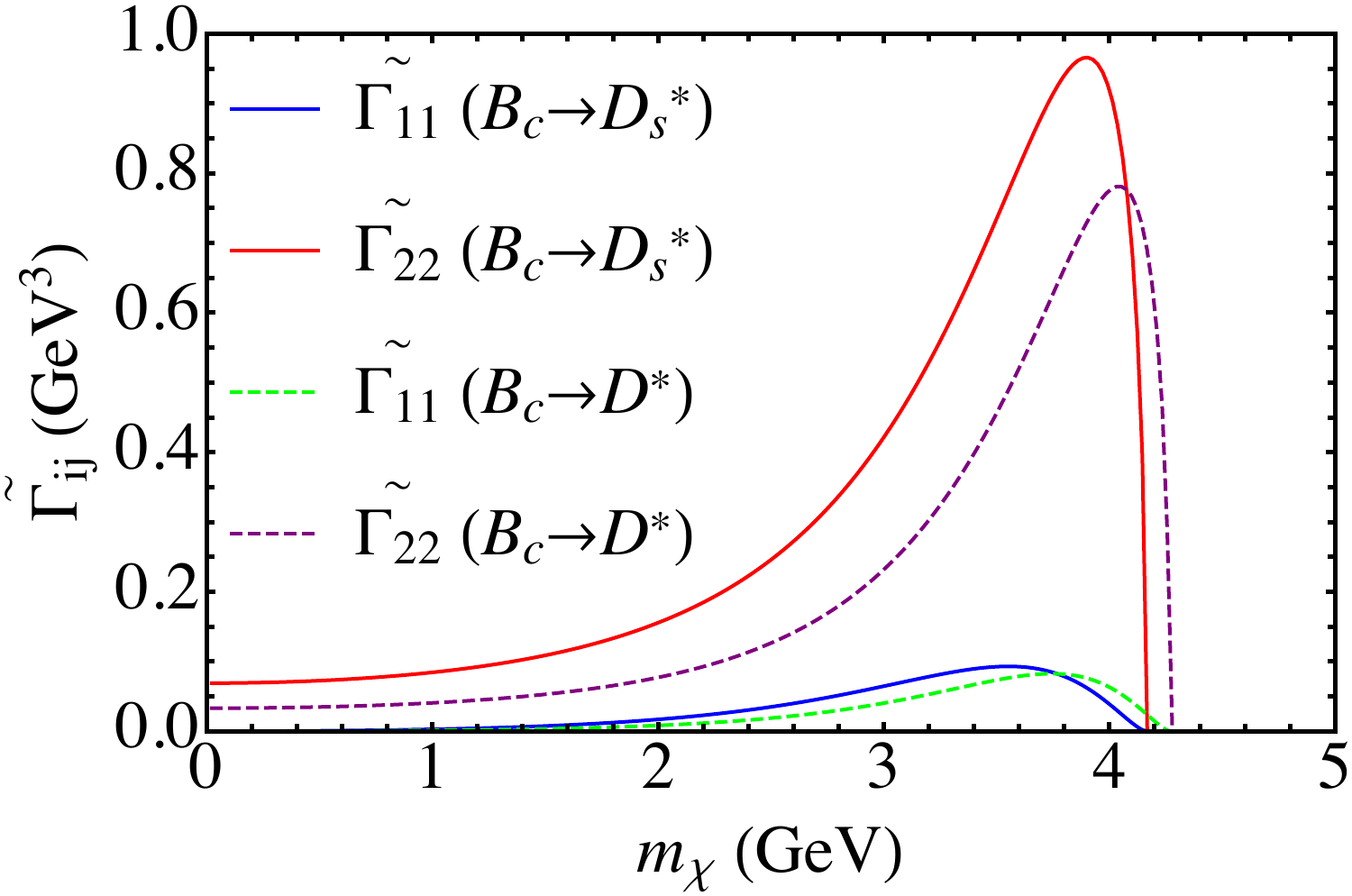}}
	\caption{$\widetilde\Gamma_{ij}$ in $0^-\to 1^-$ meson decays with invisible vector.}
	\label{width-V34}
\end{figure}

In $M^- \to M_f^{*-} V$ processes, the decay width has the form
\begin{equation}
\begin{aligned}
\Gamma(M\to M^*_fV)=& \frac{g_{_{V1}}^2}{16\pi M^3 } \lambda^{1/2}(M^2,M_f^{*2},m_{_V}^2) \bigg\{ \langle M_f^{*-}|(\bar q_{_f} \gamma_{\nu}q)|M^-\rangle^*\langle M_f^{*-}|(\bar q_{_f} \gamma_{\mu}q)|M^-\rangle P_{_V}^\mu  P_{_V}^\nu \\
&-m_{_V}^2 \langle M_f^{*-}|(\bar q_{_f} \gamma_{\mu} q)|M^-\rangle^*\langle M_f^{*-}|(\bar q_{_f} \gamma^{\mu} q)|M^-\rangle \bigg\} \\
+&\frac{g_{_{V2}}^2}{16\pi M^3 } \lambda^{1/2}(M^2,M_f^{*2},m_{_V}^2) \bigg\{ \langle M_f^{*-}|(\bar q_{_f} \gamma_{\nu}\gamma^5 q)|M^-\rangle^*\langle M_f^{*-}|(\bar q_{_f} \gamma_{\mu}\gamma^5 q)|M^-\rangle P_{_V}^\mu P_{_V}^\nu\\
&-m_{_V}^2 \langle M_f^{*-}|(\bar q_{_f} \gamma_{\mu}\gamma^5 q)|M^-\rangle^*\langle M_f^{*-}|(\bar q_{_f} \gamma^{\mu}\gamma^5 q)|M^-\rangle \bigg\}. 
\label{eq11}
\end{aligned}
\end{equation}
The hadronic transition matrix elements can be expressed as the functions of form factors in Eq.~(\ref{eq7}). In Fig.~\ref{width-V34}, the results of $\widetilde\Gamma_{ij}$s as a function of $m_{_V}$ are shown. 
$\widetilde\Gamma_{22}$ has the same shape as that in $0^-\to0^-$ modes above in Fig.~\ref{width-V12}. $\widetilde\Gamma_{11}$ starts from zero because it is proportional to $ m_{_V}^2$. There is no term like $\widetilde\Gamma_{12}$ since the interference term can be proved to be zero.

\begin{figure}[h]
	\centering
	\subfigure[$B^-\rightarrow K^{*-}V$]{
		\includegraphics[width=0.46\textwidth]{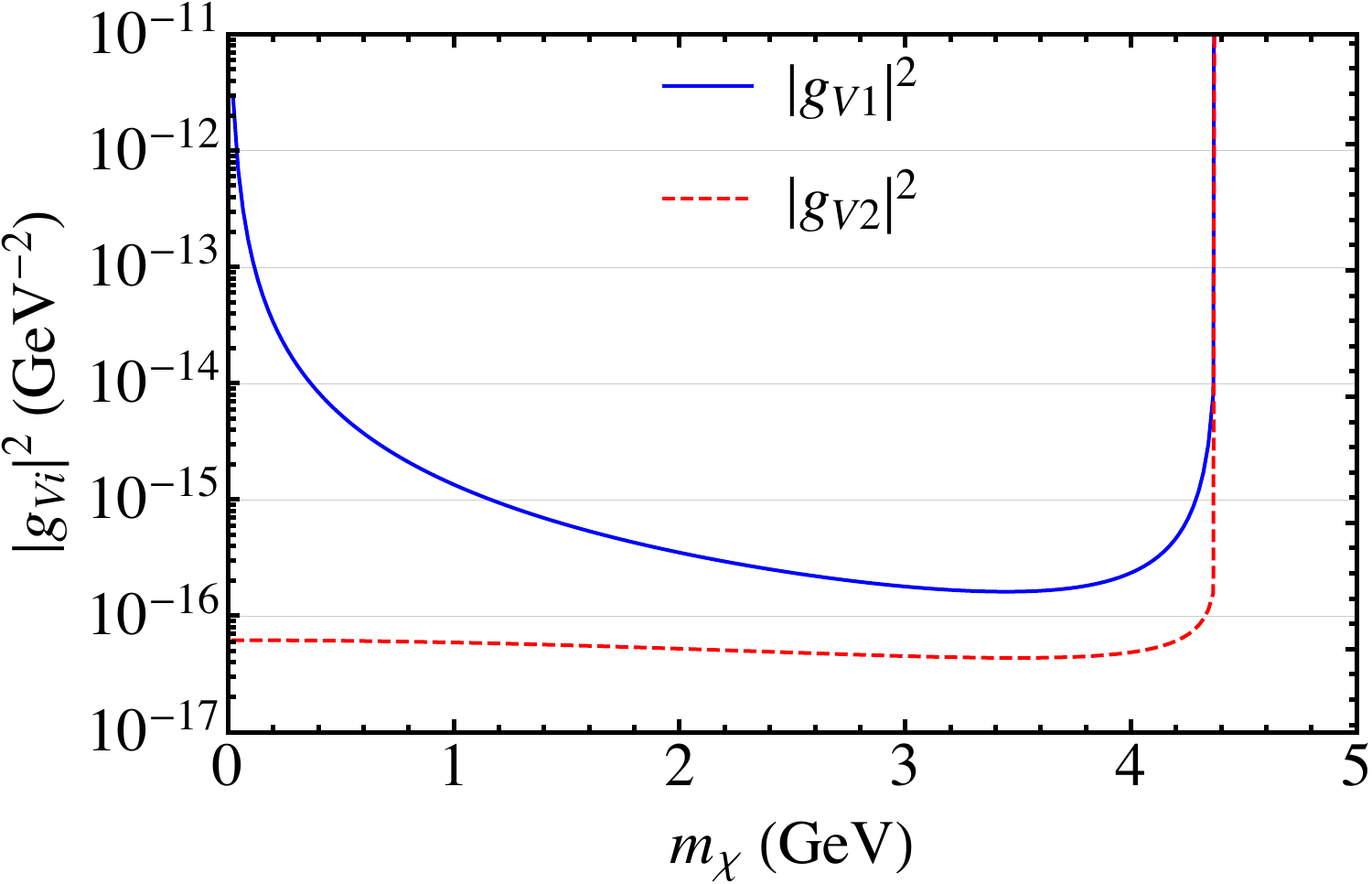}}
	\hspace{2em}
	\subfigure[$B^-\rightarrow \rho^-V$]{
		\includegraphics[width=0.46\textwidth]{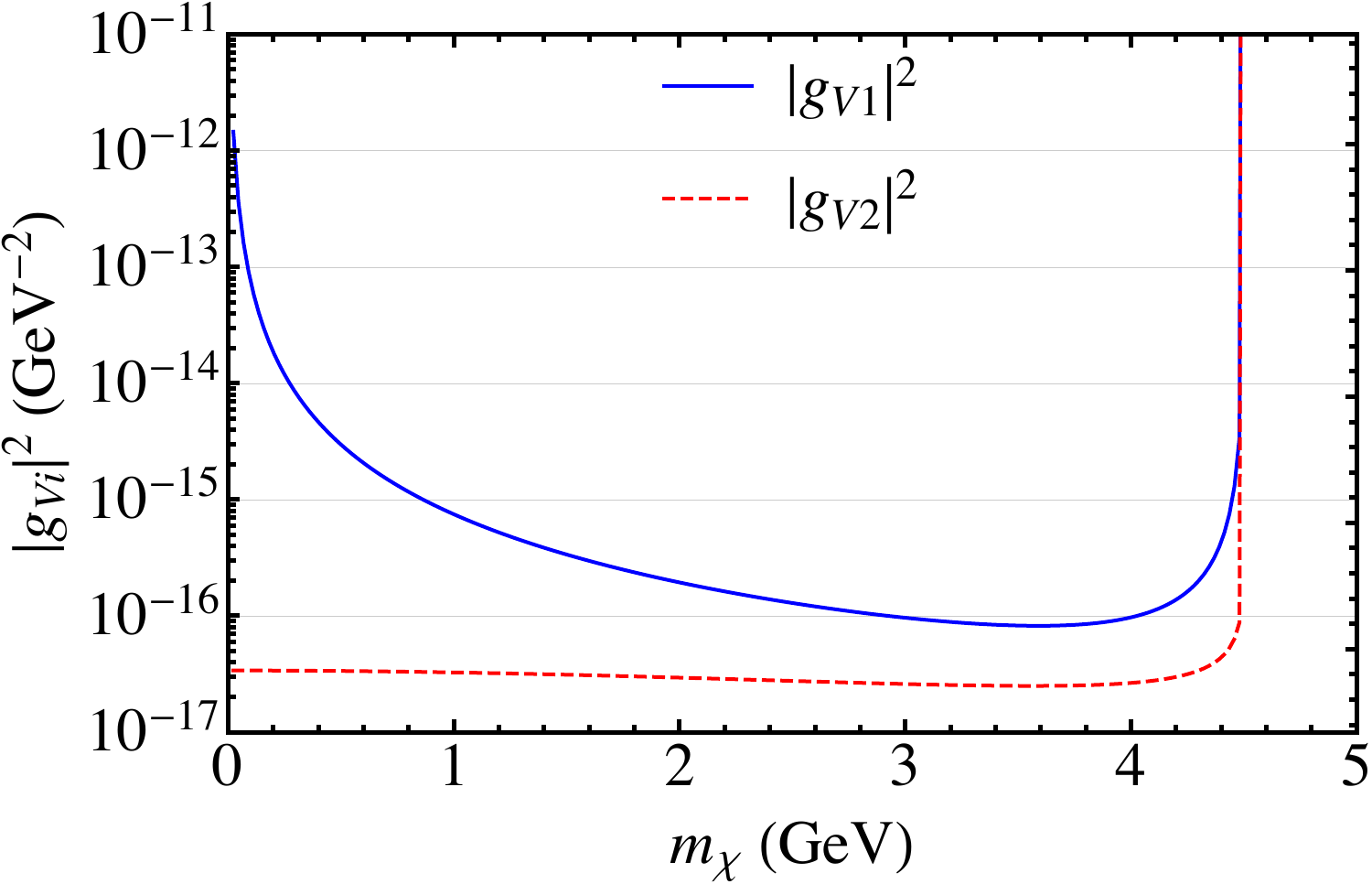}}
	\caption{Upper limits of $g_{_{Vi}}$ from $B$ meson $0^-\to 1^-$ decays with invisible vector.}
	\label{gv12}
\end{figure}

\begin{figure}[h]
	\centering
	\subfigure[$B_c^-\rightarrow D_s^{*-}V$]{
		\includegraphics[width=0.44\textwidth]{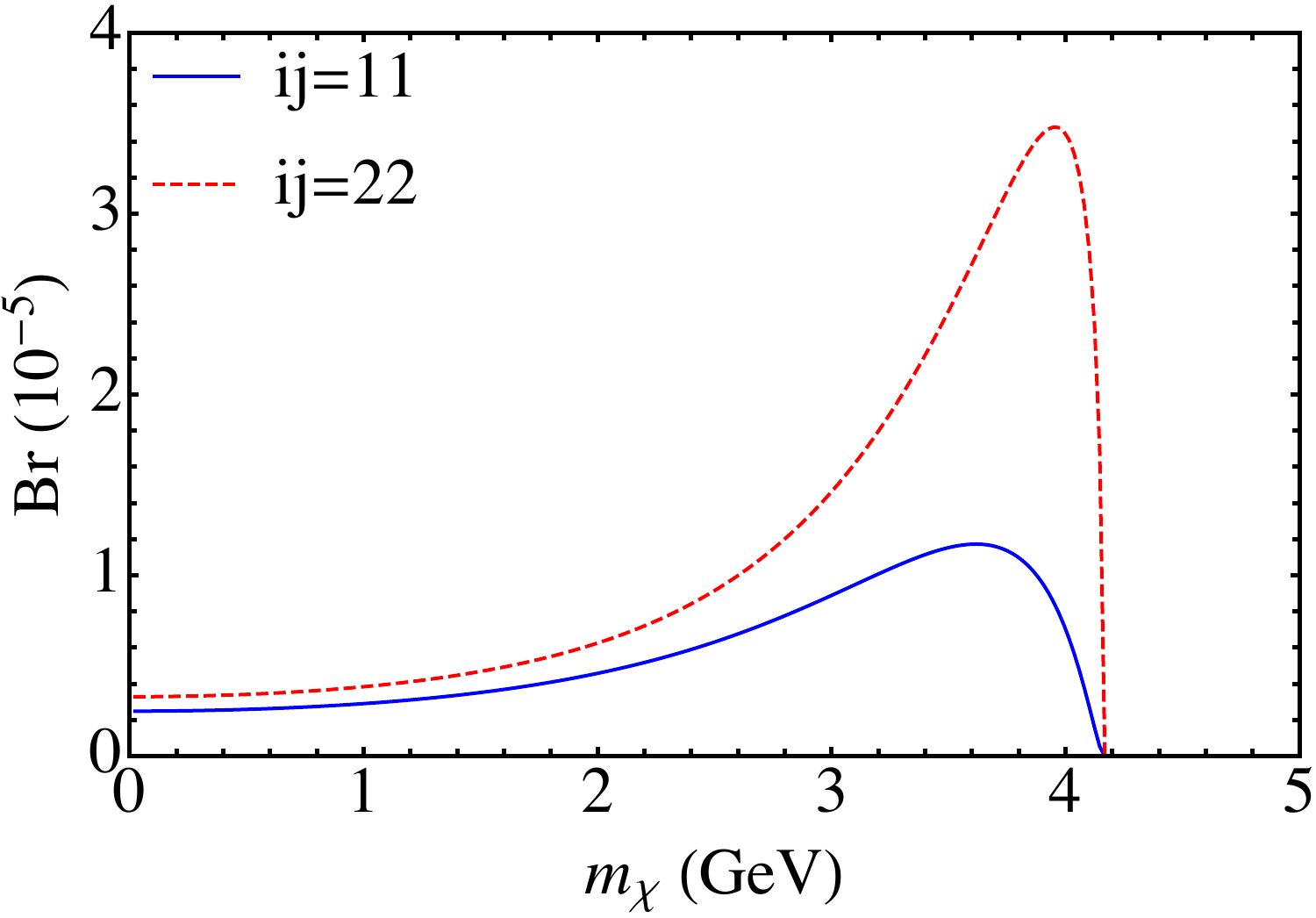}}
	\hspace{2em}
	\subfigure[$B_c^-\rightarrow D^{*-}V$]{
		\includegraphics[width=0.45\textwidth]{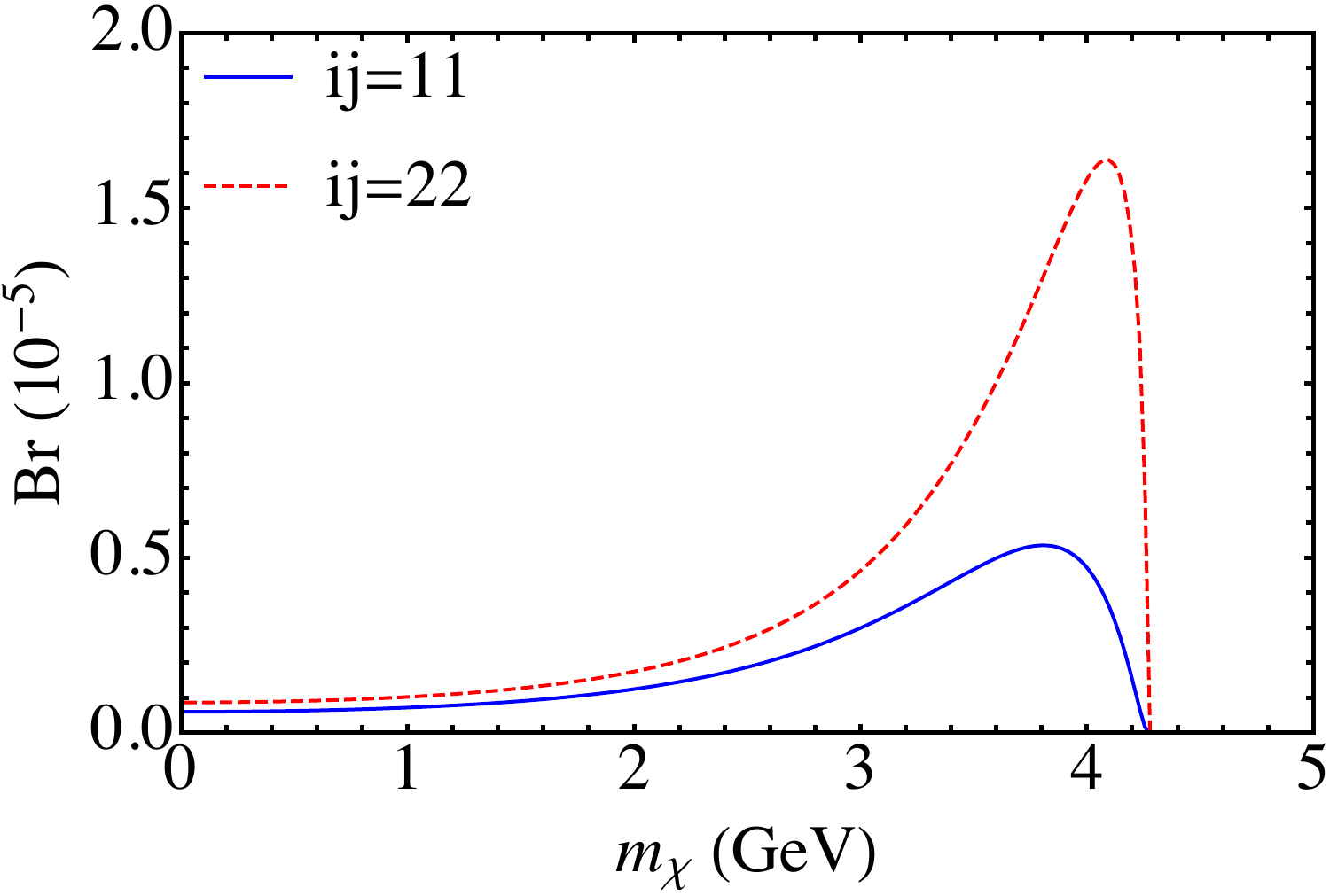}}
	\caption{Branching ratios of $B_c$ meson $0^-\to1^-$ decays with invisible vector.}
	\label{br-1}
\end{figure}

The upper limits of $|g_{_{Vi}}|^2$ are shown in Fig.~\ref{gv12}. The $|g_{_{V2}}|^2$ which is of the order of $10^{-17}$ changes slowly when $m_{_V} \textless M-M_f$. When $m_{_V}\to 0$, the upper limits of $|g_{_{V1}}|^2$ go to infinity. These results depend on $\widetilde\Gamma_{ij}$ in Fig.~\ref{width-V34a}.
The upper limits of the branching ratios are shown as Fig.~\ref{br-1}. 
One of the two operators is opened in turn, while assuming the other is zero. The blue solid line and red dashed line represent the contribution from $\widetilde\Gamma_{11}$ and $\widetilde\Gamma_{22}$, respectively. As there is no interference term, the upper limit of the branching ratio is the larger one of these lines, namely, the red dashed line.

\section{Conclusion}
We have studied the light invisible bosonic particles via FCNC processes of $B$ and $B_c$ meson. The mass is considered to be less than a few GeV. Both scalar and vector cases are considered. The effective Lagrangian is introduced to describe the coupling between quarks and the dark boson. The effective coupling constants are constrained by the experimental results for the $B$ decays with missing energy. Then the upper limits of the branching fractions of the $B_c\to M_f^{(*)}\chi$ channels are calculated. When the final meson is pseudoscalar $D_{(s)}$, the largest value of the upper limits is of the order of $10^{-6}$. For the final vector meson $D_{(s)}^*$, the $\mathcal {BR}$ is of the order of $10^{-5}$. The most likely area for finding the dark boson is near $m_\chi\approx 3.5-4$~GeV. As much more $B_c$ events will be generated in the near future, we hope future experiments can make new discoveries through such processes or set more stringent constraints for them. 

\section{Acknowledgments}

This work was supported in part by the National Natural Science Foundation of China (NSFC) under Grant No.~12075073. We also thank the HEPC Studio at Physics School of Harbin Institute of Technology for access to high performance computing resources through INSPUR-HPC@hepc.hit.edu.cn

\bibliography{reference}

\begin{thebibliography}{59}%
\makeatletter
\providecommand \@ifxundefined [1]{%
 \@ifx{#1\undefined}
}%
\providecommand \@ifnum [1]{%
 \ifnum #1\expandafter \@firstoftwo
 \else \expandafter \@secondoftwo
 \fi
}%
\providecommand \@ifx [1]{%
 \ifx #1\expandafter \@firstoftwo
 \else \expandafter \@secondoftwo
 \fi
}%
\providecommand \natexlab [1]{#1}%
\providecommand \enquote  [1]{``#1''}%
\providecommand \bibnamefont  [1]{#1}%
\providecommand \bibfnamefont [1]{#1}%
\providecommand \citenamefont [1]{#1}%
\providecommand \href@noop [0]{\@secondoftwo}%
\providecommand \href [0]{\begingroup \@sanitize@url \@href}%
\providecommand \@href[1]{\@@startlink{#1}\@@href}%
\providecommand \@@href[1]{\endgroup#1\@@endlink}%
\providecommand \@sanitize@url [0]{\catcode `\\12\catcode `\$12\catcode
  `\&12\catcode `\#12\catcode `\^12\catcode `\_12\catcode `\%12\relax}%
\providecommand \@@startlink[1]{}%
\providecommand \@@endlink[0]{}%
\providecommand \url  [0]{\begingroup\@sanitize@url \@url }%
\providecommand \@url [1]{\endgroup\@href {#1}{\urlprefix }}%
\providecommand \urlprefix  [0]{URL }%
\providecommand \Eprint [0]{\href }%
\providecommand \doibase [0]{http://dx.doi.org/}%
\providecommand \selectlanguage [0]{\@gobble}%
\providecommand \bibinfo  [0]{\@secondoftwo}%
\providecommand \bibfield  [0]{\@secondoftwo}%
\providecommand \translation [1]{[#1]}%
\providecommand \BibitemOpen [0]{}%
\providecommand \bibitemStop [0]{}%
\providecommand \bibitemNoStop [0]{.\EOS\space}%
\providecommand \EOS [0]{\spacefactor3000\relax}%
\providecommand \BibitemShut  [1]{\csname bibitem#1\endcsname}%
\let\auto@bib@innerbib\@empty
\bibitem [{\citenamefont {Bernstein}\ \emph {et~al.}(1985)\citenamefont
  {Bernstein}, \citenamefont {Brown},\ and\ \citenamefont
  {Feinberg}}]{Bernstein:1985th}%
  \BibitemOpen
  \bibfield  {author} {\bibinfo {author} {\bibfnamefont {J.}~\bibnamefont
  {Bernstein}}, \bibinfo {author} {\bibfnamefont {L.~S.}\ \bibnamefont
  {Brown}}, \ and\ \bibinfo {author} {\bibfnamefont {G.}~\bibnamefont
  {Feinberg}},\ }\href {\doibase 10.1103/PhysRevD.32.3261} {\bibfield
  {journal} {\bibinfo  {journal} {Phys. Rev. D}\ }\textbf {\bibinfo {volume}
  {32}},\ \bibinfo {pages} {3261} (\bibinfo {year} {1985})}\BibitemShut
  {NoStop}%
\bibitem [{\citenamefont {Srednicki}\ \emph {et~al.}(1988)\citenamefont
  {Srednicki}, \citenamefont {Watkins},\ and\ \citenamefont
  {Olive}}]{Srednicki:1988ce}%
  \BibitemOpen
  \bibfield  {author} {\bibinfo {author} {\bibfnamefont {M.}~\bibnamefont
  {Srednicki}}, \bibinfo {author} {\bibfnamefont {R.}~\bibnamefont {Watkins}},
  \ and\ \bibinfo {author} {\bibfnamefont {K.~A.}\ \bibnamefont {Olive}},\
  }\href {\doibase 10.1016/0550-3213(88)90099-5} {\bibfield  {journal}
  {\bibinfo  {journal} {Nucl. Phys. B}\ }\textbf {\bibinfo {volume} {310}},\
  \bibinfo {pages} {693} (\bibinfo {year} {1988})}\BibitemShut {NoStop}%
\bibitem [{\citenamefont {Izaguirre}\ \emph {et~al.}(2015)\citenamefont
  {Izaguirre}, \citenamefont {Krnjaic}, \citenamefont {Schuster},\ and\
  \citenamefont {Toro}}]{Izaguirre:2015yja}%
  \BibitemOpen
  \bibfield  {author} {\bibinfo {author} {\bibfnamefont {E.}~\bibnamefont
  {Izaguirre}}, \bibinfo {author} {\bibfnamefont {G.}~\bibnamefont {Krnjaic}},
  \bibinfo {author} {\bibfnamefont {P.}~\bibnamefont {Schuster}}, \ and\
  \bibinfo {author} {\bibfnamefont {N.}~\bibnamefont {Toro}},\ }\href {\doibase
  10.1103/PhysRevLett.115.251301} {\bibfield  {journal} {\bibinfo  {journal}
  {Phys. Rev. Lett.}\ }\textbf {\bibinfo {volume} {115}},\ \bibinfo {pages}
  {251301} (\bibinfo {year} {2015})}\BibitemShut {NoStop}%
\bibitem [{\citenamefont {Bertone}\ \emph {et~al.}(2005)\citenamefont
  {Bertone}, \citenamefont {Hooper},\ and\ \citenamefont
  {Silk}}]{Bertone:2004pz}%
  \BibitemOpen
  \bibfield  {author} {\bibinfo {author} {\bibfnamefont {G.}~\bibnamefont
  {Bertone}}, \bibinfo {author} {\bibfnamefont {D.}~\bibnamefont {Hooper}}, \
  and\ \bibinfo {author} {\bibfnamefont {J.}~\bibnamefont {Silk}},\ }\href
  {\doibase 10.1016/j.physrep.2004.08.031} {\bibfield  {journal} {\bibinfo
  {journal} {Phys. Rept.}\ }\textbf {\bibinfo {volume} {405}},\ \bibinfo
  {pages} {279} (\bibinfo {year} {2005})}\BibitemShut {NoStop}%
\bibitem [{\citenamefont {Komatsu}\ \emph {et~al.}(2009)\citenamefont {Komatsu}
  \emph {et~al.}}]{Komatsu:2008hk}%
  \BibitemOpen
  \bibfield  {author} {\bibinfo {author} {\bibfnamefont {E.}~\bibnamefont
  {Komatsu}} \emph {et~al.} (\bibinfo {collaboration} {WMAP Collaboration}),\
  }\href {\doibase 10.1088/0067-0049/180/2/330} {\bibfield  {journal} {\bibinfo
   {journal} {Astrophys. J. Suppl.}\ }\textbf {\bibinfo {volume} {180}},\
  \bibinfo {pages} {330} (\bibinfo {year} {2009})}\BibitemShut {NoStop}%
\bibitem [{\citenamefont {Akerib}\ \emph {et~al.}(2017)\citenamefont {Akerib}
  \emph {et~al.}}]{Akerib:2016vxi}%
  \BibitemOpen
  \bibfield  {author} {\bibinfo {author} {\bibfnamefont {D.}~\bibnamefont
  {Akerib}} \emph {et~al.} (\bibinfo {collaboration} {LUX Collaboration}),\
  }\href {\doibase 10.1103/PhysRevLett.118.021303} {\bibfield  {journal}
  {\bibinfo  {journal} {Phys. Rev. Lett.}\ }\textbf {\bibinfo {volume} {118}},\
  \bibinfo {pages} {021303} (\bibinfo {year} {2017})}\BibitemShut {NoStop}%
\bibitem [{\citenamefont {Cui}\ \emph {et~al.}(2017)\citenamefont {Cui} \emph
  {et~al.}}]{Cui:2017nnn}%
  \BibitemOpen
  \bibfield  {author} {\bibinfo {author} {\bibfnamefont {X.}~\bibnamefont
  {Cui}} \emph {et~al.} (\bibinfo {collaboration} {PandaX-II Collaboration}),\
  }\href {\doibase 10.1103/PhysRevLett.119.181302} {\bibfield  {journal}
  {\bibinfo  {journal} {Phys. Rev. Lett.}\ }\textbf {\bibinfo {volume} {119}},\
  \bibinfo {pages} {181302} (\bibinfo {year} {2017})}\BibitemShut {NoStop}%
\bibitem [{\citenamefont {Aprile}\ \emph {et~al.}(2018)\citenamefont {Aprile}
  \emph {et~al.}}]{Aprile:2018dbl}%
  \BibitemOpen
  \bibfield  {author} {\bibinfo {author} {\bibfnamefont {E.}~\bibnamefont
  {Aprile}} \emph {et~al.} (\bibinfo {collaboration} {XENON Collaboration}),\
  }\href {\doibase 10.1103/PhysRevLett.121.111302} {\bibfield  {journal}
  {\bibinfo  {journal} {Phys. Rev. Lett.}\ }\textbf {\bibinfo {volume} {121}},\
  \bibinfo {pages} {111302} (\bibinfo {year} {2018})}\BibitemShut {NoStop}%
\bibitem [{\citenamefont {Gligorov}\ \emph {et~al.}(2018)\citenamefont
  {Gligorov}, \citenamefont {Knapen}, \citenamefont {Papucci},\ and\
  \citenamefont {Robinson}}]{Gligorov:2017nwh}%
  \BibitemOpen
  \bibfield  {author} {\bibinfo {author} {\bibfnamefont {V.~V.}\ \bibnamefont
  {Gligorov}}, \bibinfo {author} {\bibfnamefont {S.}~\bibnamefont {Knapen}},
  \bibinfo {author} {\bibfnamefont {M.}~\bibnamefont {Papucci}}, \ and\
  \bibinfo {author} {\bibfnamefont {D.~J.}\ \bibnamefont {Robinson}},\ }\href
  {\doibase 10.1103/PhysRevD.97.015023} {\bibfield  {journal} {\bibinfo
  {journal} {Phys. Rev. D}\ }\textbf {\bibinfo {volume} {97}},\ \bibinfo
  {pages} {015023} (\bibinfo {year} {2018})}\BibitemShut {NoStop}%
\bibitem [{\citenamefont {Lee}\ and\ \citenamefont
  {Weinberg}(1977)}]{PhysRevLett.39.165}%
  \BibitemOpen
  \bibfield  {author} {\bibinfo {author} {\bibfnamefont {B.~W.}\ \bibnamefont
  {Lee}}\ and\ \bibinfo {author} {\bibfnamefont {S.}~\bibnamefont {Weinberg}},\
  }\href {\doibase 10.1103/PhysRevLett.39.165} {\bibfield  {journal} {\bibinfo
  {journal} {Phys. Rev. Lett.}\ }\textbf {\bibinfo {volume} {39}},\ \bibinfo
  {pages} {165} (\bibinfo {year} {1977})}\BibitemShut {NoStop}%
\bibitem [{\citenamefont {Pospelov}\ \emph
  {et~al.}(2008{\natexlab{a}})\citenamefont {Pospelov}, \citenamefont {Ritz},\
  and\ \citenamefont {Voloshin}}]{Pospelov:2007mp}%
  \BibitemOpen
  \bibfield  {author} {\bibinfo {author} {\bibfnamefont {M.}~\bibnamefont
  {Pospelov}}, \bibinfo {author} {\bibfnamefont {A.}~\bibnamefont {Ritz}}, \
  and\ \bibinfo {author} {\bibfnamefont {M.~B.}\ \bibnamefont {Voloshin}},\
  }\href {\doibase 10.1016/j.physletb.2008.02.052} {\bibfield  {journal}
  {\bibinfo  {journal} {Phys. Lett. B}\ }\textbf {\bibinfo {volume} {662}},\
  \bibinfo {pages} {53} (\bibinfo {year} {2008}{\natexlab{a}})}\BibitemShut
  {NoStop}%
\bibitem [{\citenamefont {Hooper}\ and\ \citenamefont
  {Zurek}(2008)}]{Hooper:2008im}%
  \BibitemOpen
  \bibfield  {author} {\bibinfo {author} {\bibfnamefont {D.}~\bibnamefont
  {Hooper}}\ and\ \bibinfo {author} {\bibfnamefont {K.~M.}\ \bibnamefont
  {Zurek}},\ }\href {\doibase 10.1103/PhysRevD.77.087302} {\bibfield  {journal}
  {\bibinfo  {journal} {Phys. Rev. D}\ }\textbf {\bibinfo {volume} {77}},\
  \bibinfo {pages} {087302} (\bibinfo {year} {2008})}\BibitemShut {NoStop}%
\bibitem [{\citenamefont {McDonald}(2002)}]{McDonald:2001vt}%
  \BibitemOpen
  \bibfield  {author} {\bibinfo {author} {\bibfnamefont {J.}~\bibnamefont
  {McDonald}},\ }\href {\doibase 10.1103/PhysRevLett.88.091304} {\bibfield
  {journal} {\bibinfo  {journal} {Phys. Rev. Lett.}\ }\textbf {\bibinfo
  {volume} {88}},\ \bibinfo {pages} {091304} (\bibinfo {year}
  {2002})}\BibitemShut {NoStop}%
\bibitem [{\citenamefont {Hall}\ \emph {et~al.}()\citenamefont {Hall},
  \citenamefont {Jedamzik}, \citenamefont {March-Russell},\ and\ \citenamefont
  {West}}]{Hall:2009bx}%
  \BibitemOpen
  \bibfield  {author} {\bibinfo {author} {\bibfnamefont {L.~J.}\ \bibnamefont
  {Hall}}, \bibinfo {author} {\bibfnamefont {K.}~\bibnamefont {Jedamzik}},
  \bibinfo {author} {\bibfnamefont {J.}~\bibnamefont {March-Russell}}, \ and\
  \bibinfo {author} {\bibfnamefont {S.~M.}\ \bibnamefont {West}},\ }\href
  {\doibase 10.1007/JHEP03(2010)080} {\bibfield  {journal} {\bibinfo  {journal}
  {JHEP}\ }\textbf {\bibinfo {volume} {03}},\ \bibinfo {pages}
  {080}}\BibitemShut {NoStop}%
\bibitem [{\citenamefont {Bernal}\ \emph {et~al.}(2017)\citenamefont {Bernal},
  \citenamefont {Heikinheimo}, \citenamefont {Tenkanen}, \citenamefont
  {Tuominen},\ and\ \citenamefont {Vaskonen}}]{Bernal:2017kxu}%
  \BibitemOpen
  \bibfield  {author} {\bibinfo {author} {\bibfnamefont {N.}~\bibnamefont
  {Bernal}}, \bibinfo {author} {\bibfnamefont {M.}~\bibnamefont {Heikinheimo}},
  \bibinfo {author} {\bibfnamefont {T.}~\bibnamefont {Tenkanen}}, \bibinfo
  {author} {\bibfnamefont {K.}~\bibnamefont {Tuominen}}, \ and\ \bibinfo
  {author} {\bibfnamefont {V.}~\bibnamefont {Vaskonen}},\ }\href {\doibase
  10.1142/S0217751X1730023X} {\bibfield  {journal} {\bibinfo  {journal} {Int.
  J. Mod. Phys. A}\ }\textbf {\bibinfo {volume} {32}},\ \bibinfo {pages}
  {1730023} (\bibinfo {year} {2017})}\BibitemShut {NoStop}%
\bibitem [{\citenamefont {del Amo~Sanchez}\ \emph {et~al.}(2010)\citenamefont
  {del Amo~Sanchez} \emph {et~al.}}]{delAmoSanchez:2010bk}%
  \BibitemOpen
  \bibfield  {author} {\bibinfo {author} {\bibfnamefont {P.}~\bibnamefont {del
  Amo~Sanchez}} \emph {et~al.} (\bibinfo {collaboration} {BaBar
  Collaboration}),\ }\href {\doibase 10.1103/PhysRevD.82.112002} {\bibfield
  {journal} {\bibinfo  {journal} {Phys. Rev. D}\ }\textbf {\bibinfo {volume}
  {82}},\ \bibinfo {pages} {112002} (\bibinfo {year} {2010})}\BibitemShut
  {NoStop}%
\bibitem [{\citenamefont {Aubert}\ \emph {et~al.}(2008)\citenamefont {Aubert}
  \emph {et~al.}}]{Aubert:2008am}%
  \BibitemOpen
  \bibfield  {author} {\bibinfo {author} {\bibfnamefont {B.}~\bibnamefont
  {Aubert}} \emph {et~al.} (\bibinfo {collaboration} {BaBar Collaboration}),\
  }\href {\doibase 10.1103/PhysRevD.78.072007} {\bibfield  {journal} {\bibinfo
  {journal} {Phys. Rev. D}\ }\textbf {\bibinfo {volume} {78}},\ \bibinfo
  {pages} {072007} (\bibinfo {year} {2008})}\BibitemShut {NoStop}%
\bibitem [{\citenamefont {Chen}\ \emph {et~al.}(2007)\citenamefont {Chen} \emph
  {et~al.}}]{Chen:2007zk}%
  \BibitemOpen
  \bibfield  {author} {\bibinfo {author} {\bibfnamefont {K.~F.}\ \bibnamefont
  {Chen}} \emph {et~al.} (\bibinfo {collaboration} {Belle Collaboration}),\
  }\href {\doibase 10.1103/PhysRevLett.99.221802} {\bibfield  {journal}
  {\bibinfo  {journal} {Phys. Rev. Lett.}\ }\textbf {\bibinfo {volume} {99}},\
  \bibinfo {pages} {221802} (\bibinfo {year} {2007})}\BibitemShut {NoStop}%
\bibitem [{\citenamefont {Grygier}\ \emph {et~al.}(2017)\citenamefont {Grygier}
  \emph {et~al.}}]{Grygier:2017tzo}%
  \BibitemOpen
  \bibfield  {author} {\bibinfo {author} {\bibfnamefont {J.}~\bibnamefont
  {Grygier}} \emph {et~al.} (\bibinfo {collaboration} {Belle Collaboration}),\
  }\href {\doibase 10.1103/PhysRevD.96.091101} {\bibfield  {journal} {\bibinfo
  {journal} {Phys. Rev. D}\ }\textbf {\bibinfo {volume} {96}},\ \bibinfo
  {pages} {091101} (\bibinfo {year} {2017})}\BibitemShut {NoStop}%
\bibitem [{\citenamefont {Lai}\ \emph {et~al.}(2017)\citenamefont {Lai} \emph
  {et~al.}}]{Lai:2016uvj}%
  \BibitemOpen
  \bibfield  {author} {\bibinfo {author} {\bibfnamefont {Y.~T.}\ \bibnamefont
  {Lai}} \emph {et~al.} (\bibinfo {collaboration} {Belle Collaboration}),\
  }\href {\doibase 10.1103/PhysRevD.95.011102} {\bibfield  {journal} {\bibinfo
  {journal} {Phys. Rev. D}\ }\textbf {\bibinfo {volume} {95}},\ \bibinfo
  {pages} {011102} (\bibinfo {year} {2017})}\BibitemShut {NoStop}%
\bibitem [{\citenamefont {Bird}\ \emph {et~al.}(2004)\citenamefont {Bird},
  \citenamefont {Jackson}, \citenamefont {Kowalewski},\ and\ \citenamefont
  {Pospelov}}]{Bird:2004ts}%
  \BibitemOpen
  \bibfield  {author} {\bibinfo {author} {\bibfnamefont {C.}~\bibnamefont
  {Bird}}, \bibinfo {author} {\bibfnamefont {P.}~\bibnamefont {Jackson}},
  \bibinfo {author} {\bibfnamefont {R.~V.}\ \bibnamefont {Kowalewski}}, \ and\
  \bibinfo {author} {\bibfnamefont {M.}~\bibnamefont {Pospelov}},\ }\href
  {\doibase 10.1103/PhysRevLett.93.201803} {\bibfield  {journal} {\bibinfo
  {journal} {Phys. Rev. Lett.}\ }\textbf {\bibinfo {volume} {93}},\ \bibinfo
  {pages} {201803} (\bibinfo {year} {2004})}\BibitemShut {NoStop}%
\bibitem [{\citenamefont {Bird}\ \emph {et~al.}(2006)\citenamefont {Bird},
  \citenamefont {Kowalewski},\ and\ \citenamefont {Pospelov}}]{Bird:2006jd}%
  \BibitemOpen
  \bibfield  {author} {\bibinfo {author} {\bibfnamefont {C.}~\bibnamefont
  {Bird}}, \bibinfo {author} {\bibfnamefont {R.~V.}\ \bibnamefont
  {Kowalewski}}, \ and\ \bibinfo {author} {\bibfnamefont {M.}~\bibnamefont
  {Pospelov}},\ }\href {\doibase 10.1142/S0217732306019852} {\bibfield
  {journal} {\bibinfo  {journal} {Mod. Phys. Lett. A}\ }\textbf {\bibinfo
  {volume} {21}},\ \bibinfo {pages} {457} (\bibinfo {year} {2006})}\BibitemShut
  {NoStop}%
\bibitem [{\citenamefont {Badin}\ and\ \citenamefont
  {Petrov}(2010)}]{Badin:2010uh}%
  \BibitemOpen
  \bibfield  {author} {\bibinfo {author} {\bibfnamefont {A.}~\bibnamefont
  {Badin}}\ and\ \bibinfo {author} {\bibfnamefont {A.~A.}\ \bibnamefont
  {Petrov}},\ }\href {\doibase 10.1103/PhysRevD.82.034005} {\bibfield
  {journal} {\bibinfo  {journal} {Phys. Rev. D}\ }\textbf {\bibinfo {volume}
  {82}},\ \bibinfo {pages} {034005} (\bibinfo {year} {2010})}\BibitemShut
  {NoStop}%
\bibitem [{\citenamefont {Gninenko}\ and\ \citenamefont
  {Krasnikov}(2015)}]{Gninenko:2015mea}%
  \BibitemOpen
  \bibfield  {author} {\bibinfo {author} {\bibfnamefont {S.~N.}\ \bibnamefont
  {Gninenko}}\ and\ \bibinfo {author} {\bibfnamefont {N.~V.}\ \bibnamefont
  {Krasnikov}},\ }\href {\doibase 10.1103/PhysRevD.92.034009} {\bibfield
  {journal} {\bibinfo  {journal} {Phys. Rev. D}\ }\textbf {\bibinfo {volume}
  {92}},\ \bibinfo {pages} {034009} (\bibinfo {year} {2015})}\BibitemShut
  {NoStop}%
\bibitem [{\citenamefont {Barducci}\ \emph {et~al.}(2018)\citenamefont
  {Barducci}, \citenamefont {Fabbrichesi},\ and\ \citenamefont
  {Gabrielli}}]{Barducci:2018rlx}%
  \BibitemOpen
  \bibfield  {author} {\bibinfo {author} {\bibfnamefont {D.}~\bibnamefont
  {Barducci}}, \bibinfo {author} {\bibfnamefont {M.}~\bibnamefont
  {Fabbrichesi}}, \ and\ \bibinfo {author} {\bibfnamefont {E.}~\bibnamefont
  {Gabrielli}},\ }\href {\doibase 10.1103/PhysRevD.98.035049} {\bibfield
  {journal} {\bibinfo  {journal} {Phys. Rev. D}\ }\textbf {\bibinfo {volume}
  {98}},\ \bibinfo {pages} {035049} (\bibinfo {year} {2018})}\BibitemShut
  {NoStop}%
\bibitem [{\citenamefont {Kamenik}\ and\ \citenamefont
  {Smith}(2012)}]{Kamenik:2011vy}%
  \BibitemOpen
  \bibfield  {author} {\bibinfo {author} {\bibfnamefont {J.~F.}\ \bibnamefont
  {Kamenik}}\ and\ \bibinfo {author} {\bibfnamefont {C.}~\bibnamefont
  {Smith}},\ }\href {\doibase 10.1007/JHEP03(2012)090} {\bibfield  {journal}
  {\bibinfo  {journal} {JHEP}\ }\textbf {\bibinfo {volume} {03}},\ \bibinfo
  {pages} {090} (\bibinfo {year} {2012})}\BibitemShut {NoStop}%
\bibitem [{\citenamefont {Bertuzzo}\ \emph {et~al.}(2017)\citenamefont
  {Bertuzzo}, \citenamefont {Caniu~Barros},\ and\ \citenamefont {Grilli~di
  Cortona}}]{Bertuzzo:2017lwt}%
  \BibitemOpen
  \bibfield  {author} {\bibinfo {author} {\bibfnamefont {E.}~\bibnamefont
  {Bertuzzo}}, \bibinfo {author} {\bibfnamefont {C.~J.}\ \bibnamefont
  {Caniu~Barros}}, \ and\ \bibinfo {author} {\bibfnamefont {G.}~\bibnamefont
  {Grilli~di Cortona}},\ }\href {\doibase 10.1007/JHEP09(2017)116} {\bibfield
  {journal} {\bibinfo  {journal} {JHEP}\ }\textbf {\bibinfo {volume} {09}},\
  \bibinfo {pages} {116} (\bibinfo {year} {2017})}\BibitemShut {NoStop}%
\bibitem [{\citenamefont {Aaltonen}\ \emph {et~al.}(2016)\citenamefont
  {Aaltonen} \emph {et~al.}}]{Aaltonen:2016dra}%
  \BibitemOpen
  \bibfield  {author} {\bibinfo {author} {\bibfnamefont {T.~A.}\ \bibnamefont
  {Aaltonen}} \emph {et~al.} (\bibinfo {collaboration} {CDF Collaboration}),\
  }\href {\doibase 10.1103/PhysRevD.93.052001} {\bibfield  {journal} {\bibinfo
  {journal} {Phys. Rev. D}\ }\textbf {\bibinfo {volume} {93}},\ \bibinfo
  {pages} {052001} (\bibinfo {year} {2016})}\BibitemShut {NoStop}%
\bibitem [{\citenamefont {Burdin}(2016)}]{Burdin:2016rzf}%
  \BibitemOpen
  \bibfield  {author} {\bibinfo {author} {\bibfnamefont {S.}~\bibnamefont
  {Burdin}} (\bibinfo {collaboration} {ATLAS Collaboration}),\ }\href {\doibase
  10.1063/1.4949386} {\bibfield  {journal} {\bibinfo  {journal} {AIP Conf.
  Proc.}\ }\textbf {\bibinfo {volume} {1735}},\ \bibinfo {pages} {030003}
  (\bibinfo {year} {2016})}\BibitemShut {NoStop}%
\bibitem [{\citenamefont {Berezhnoy}\ \emph {et~al.}(2019)\citenamefont
  {Berezhnoy}, \citenamefont {Belov}, \citenamefont {Likhoded},\ and\
  \citenamefont {Luhinsky}}]{Berezhnoy:2019yei}%
  \BibitemOpen
  \bibfield  {author} {\bibinfo {author} {\bibfnamefont {A.}~\bibnamefont
  {Berezhnoy}}, \bibinfo {author} {\bibfnamefont {I.}~\bibnamefont {Belov}},
  \bibinfo {author} {\bibfnamefont {A.}~\bibnamefont {Likhoded}}, \ and\
  \bibinfo {author} {\bibfnamefont {A.}~\bibnamefont {Luhinsky}},\ }\href
  {\doibase 10.1142/S0217732319503310} {\bibfield  {journal} {\bibinfo
  {journal} {Mod. Phys. Lett. A}\ }\textbf {\bibinfo {volume} {34}},\ \bibinfo
  {pages} {1950331} (\bibinfo {year} {2019})}\BibitemShut {NoStop}%
\bibitem [{\citenamefont {Aaij}\ \emph {et~al.}(2019)\citenamefont {Aaij} \emph
  {et~al.}}]{Aaij:2019ths}%
  \BibitemOpen
  \bibfield  {author} {\bibinfo {author} {\bibfnamefont {R.}~\bibnamefont
  {Aaij}} \emph {et~al.} (\bibinfo {collaboration} {LHCb Collaboration}),\
  }\href {\doibase 10.1103/PhysRevD.100.112006} {\bibfield  {journal} {\bibinfo
   {journal} {Phys. Rev. D}\ }\textbf {\bibinfo {volume} {100}},\ \bibinfo
  {pages} {112006} (\bibinfo {year} {2019})}\BibitemShut {NoStop}%
\bibitem [{\citenamefont {Pospelov}\ \emph
  {et~al.}(2008{\natexlab{b}})\citenamefont {Pospelov}, \citenamefont {Ritz},\
  and\ \citenamefont {Voloshin}}]{Pospelov:2008jk}%
  \BibitemOpen
  \bibfield  {author} {\bibinfo {author} {\bibfnamefont {M.}~\bibnamefont
  {Pospelov}}, \bibinfo {author} {\bibfnamefont {A.}~\bibnamefont {Ritz}}, \
  and\ \bibinfo {author} {\bibfnamefont {M.~B.}\ \bibnamefont {Voloshin}},\
  }\href {\doibase 10.1103/PhysRevD.78.115012} {\bibfield  {journal} {\bibinfo
  {journal} {Phys. Rev. D}\ }\textbf {\bibinfo {volume} {78}},\ \bibinfo
  {pages} {115012} (\bibinfo {year} {2008}{\natexlab{b}})}\BibitemShut
  {NoStop}%
\bibitem [{\citenamefont {Redondo}\ and\ \citenamefont
  {Postma}(2009)}]{Redondo:2008ec}%
  \BibitemOpen
  \bibfield  {author} {\bibinfo {author} {\bibfnamefont {J.}~\bibnamefont
  {Redondo}}\ and\ \bibinfo {author} {\bibfnamefont {M.}~\bibnamefont
  {Postma}},\ }\href {\doibase 10.1088/1475-7516/2009/02/005} {\bibfield
  {journal} {\bibinfo  {journal} {JCAP}\ }\textbf {\bibinfo {volume} {02}},\
  \bibinfo {pages} {005} (\bibinfo {year} {2009})}\BibitemShut {NoStop}%
\bibitem [{\citenamefont {Bjorken}\ \emph {et~al.}(2009)\citenamefont
  {Bjorken}, \citenamefont {Essig}, \citenamefont {Schuster},\ and\
  \citenamefont {Toro}}]{Bjorken:2009mm}%
  \BibitemOpen
  \bibfield  {author} {\bibinfo {author} {\bibfnamefont {J.~D.}\ \bibnamefont
  {Bjorken}}, \bibinfo {author} {\bibfnamefont {R.}~\bibnamefont {Essig}},
  \bibinfo {author} {\bibfnamefont {P.}~\bibnamefont {Schuster}}, \ and\
  \bibinfo {author} {\bibfnamefont {N.}~\bibnamefont {Toro}},\ }\href {\doibase
  10.1103/PhysRevD.80.075018} {\bibfield  {journal} {\bibinfo  {journal} {Phys.
  Rev. D}\ }\textbf {\bibinfo {volume} {80}},\ \bibinfo {pages} {075018}
  (\bibinfo {year} {2009})}\BibitemShut {NoStop}%
\bibitem [{\citenamefont {Diaz-Cruz}\ and\ \citenamefont
  {Ma}(2011)}]{DiazCruz:2010dc}%
  \BibitemOpen
  \bibfield  {author} {\bibinfo {author} {\bibfnamefont {J.~L.}\ \bibnamefont
  {Diaz-Cruz}}\ and\ \bibinfo {author} {\bibfnamefont {E.}~\bibnamefont {Ma}},\
  }\href {\doibase 10.1016/j.physletb.2010.11.039} {\bibfield  {journal}
  {\bibinfo  {journal} {Phys. Lett. B}\ }\textbf {\bibinfo {volume} {695}},\
  \bibinfo {pages} {264} (\bibinfo {year} {2011})}\BibitemShut {NoStop}%
\bibitem [{\citenamefont {Baek}\ \emph {et~al.}(2013)\citenamefont {Baek},
  \citenamefont {Ko}, \citenamefont {Park},\ and\ \citenamefont
  {Senaha}}]{Baek:2012se}%
  \BibitemOpen
  \bibfield  {author} {\bibinfo {author} {\bibfnamefont {S.}~\bibnamefont
  {Baek}}, \bibinfo {author} {\bibfnamefont {P.}~\bibnamefont {Ko}}, \bibinfo
  {author} {\bibfnamefont {W.-I.}\ \bibnamefont {Park}}, \ and\ \bibinfo
  {author} {\bibfnamefont {E.}~\bibnamefont {Senaha}},\ }\href {\doibase
  10.1007/JHEP05(2013)036} {\bibfield  {journal} {\bibinfo  {journal} {JHEP}\
  }\textbf {\bibinfo {volume} {05}},\ \bibinfo {pages} {036} (\bibinfo {year}
  {2013})}\BibitemShut {NoStop}%
\bibitem [{\citenamefont {Fabbrichesi}\ \emph {et~al.}(2020)\citenamefont
  {Fabbrichesi}, \citenamefont {Gabrielli},\ and\ \citenamefont
  {Lanfranchi}}]{Fabbrichesi:2020wbt}%
  \BibitemOpen
  \bibfield  {author} {\bibinfo {author} {\bibfnamefont {M.}~\bibnamefont
  {Fabbrichesi}}, \bibinfo {author} {\bibfnamefont {E.}~\bibnamefont
  {Gabrielli}}, \ and\ \bibinfo {author} {\bibfnamefont {G.}~\bibnamefont
  {Lanfranchi}},\ }\href {\doibase 10.1007/978-3-030-62519-1} {\  (\bibinfo
  {year} {2020}),\ 10.1007/978-3-030-62519-1},\ \Eprint
  {http://arxiv.org/abs/2005.01515} {arXiv:2005.01515 [hep-ph]} \BibitemShut
  {NoStop}%
\bibitem [{\citenamefont {Peccei}\ and\ \citenamefont
  {Quinn}(1977)}]{Peccei:1977hh}%
  \BibitemOpen
  \bibfield  {author} {\bibinfo {author} {\bibfnamefont {R.~D.}\ \bibnamefont
  {Peccei}}\ and\ \bibinfo {author} {\bibfnamefont {H.~R.}\ \bibnamefont
  {Quinn}},\ }\href {\doibase 10.1103/PhysRevLett.38.1440} {\bibfield
  {journal} {\bibinfo  {journal} {Phys. Rev. Lett.}\ }\textbf {\bibinfo
  {volume} {38}},\ \bibinfo {pages} {1440} (\bibinfo {year}
  {1977})}\BibitemShut {NoStop}%
\bibitem [{\citenamefont {Weinberg}(1978)}]{PhysRevLett.40.223}%
  \BibitemOpen
  \bibfield  {author} {\bibinfo {author} {\bibfnamefont {S.}~\bibnamefont
  {Weinberg}},\ }\href {\doibase 10.1103/PhysRevLett.40.223} {\bibfield
  {journal} {\bibinfo  {journal} {Phys. Rev. Lett.}\ }\textbf {\bibinfo
  {volume} {40}},\ \bibinfo {pages} {223} (\bibinfo {year} {1978})}\BibitemShut
  {NoStop}%
\bibitem [{\citenamefont {Wilczek}(1978)}]{Wilczek:1977pj}%
  \BibitemOpen
  \bibfield  {author} {\bibinfo {author} {\bibfnamefont {F.}~\bibnamefont
  {Wilczek}},\ }\href {\doibase 10.1103/PhysRevLett.40.279} {\bibfield
  {journal} {\bibinfo  {journal} {Phys. Rev. Lett.}\ }\textbf {\bibinfo
  {volume} {40}},\ \bibinfo {pages} {279} (\bibinfo {year} {1978})}\BibitemShut
  {NoStop}%
\bibitem [{\citenamefont {Batell}\ \emph {et~al.}(2011)\citenamefont {Batell},
  \citenamefont {Pospelov},\ and\ \citenamefont {Ritz}}]{Batell:2009jf}%
  \BibitemOpen
  \bibfield  {author} {\bibinfo {author} {\bibfnamefont {B.}~\bibnamefont
  {Batell}}, \bibinfo {author} {\bibfnamefont {M.}~\bibnamefont {Pospelov}}, \
  and\ \bibinfo {author} {\bibfnamefont {A.}~\bibnamefont {Ritz}},\ }\href
  {\doibase 10.1103/PhysRevD.83.054005} {\bibfield  {journal} {\bibinfo
  {journal} {Phys. Rev. D}\ }\textbf {\bibinfo {volume} {83}},\ \bibinfo
  {pages} {054005} (\bibinfo {year} {2011})}\BibitemShut {NoStop}%
\bibitem [{\citenamefont {Aditya}\ \emph {et~al.}(2012)\citenamefont {Aditya},
  \citenamefont {Healey},\ and\ \citenamefont {Petrov}}]{Aditya:2012ay}%
  \BibitemOpen
  \bibfield  {author} {\bibinfo {author} {\bibfnamefont {Y.}~\bibnamefont
  {Aditya}}, \bibinfo {author} {\bibfnamefont {K.~J.}\ \bibnamefont {Healey}},
  \ and\ \bibinfo {author} {\bibfnamefont {A.~A.}\ \bibnamefont {Petrov}},\
  }\href {\doibase 10.1016/j.physletb.2012.02.042} {\bibfield  {journal}
  {\bibinfo  {journal} {Phys. Lett. B}\ }\textbf {\bibinfo {volume} {710}},\
  \bibinfo {pages} {118} (\bibinfo {year} {2012})}\BibitemShut {NoStop}%
\bibitem [{\citenamefont {Izaguirre}\ \emph {et~al.}(2017)\citenamefont
  {Izaguirre}, \citenamefont {Lin},\ and\ \citenamefont
  {Shuve}}]{Izaguirre:2016dfi}%
  \BibitemOpen
  \bibfield  {author} {\bibinfo {author} {\bibfnamefont {E.}~\bibnamefont
  {Izaguirre}}, \bibinfo {author} {\bibfnamefont {T.}~\bibnamefont {Lin}}, \
  and\ \bibinfo {author} {\bibfnamefont {B.}~\bibnamefont {Shuve}},\ }\href
  {\doibase 10.1103/PhysRevLett.118.111802} {\bibfield  {journal} {\bibinfo
  {journal} {Phys. Rev. Lett.}\ }\textbf {\bibinfo {volume} {118}},\ \bibinfo
  {pages} {111802} (\bibinfo {year} {2017})}\BibitemShut {NoStop}%
\bibitem [{\citenamefont {O'Connell}\ \emph {et~al.}(2007)\citenamefont
  {O'Connell}, \citenamefont {Ramsey-Musolf},\ and\ \citenamefont
  {Wise}}]{OConnell:2006rsp}%
  \BibitemOpen
  \bibfield  {author} {\bibinfo {author} {\bibfnamefont {D.}~\bibnamefont
  {O'Connell}}, \bibinfo {author} {\bibfnamefont {M.~J.}\ \bibnamefont
  {Ramsey-Musolf}}, \ and\ \bibinfo {author} {\bibfnamefont {M.~B.}\
  \bibnamefont {Wise}},\ }\href {\doibase 10.1103/PhysRevD.75.037701}
  {\bibfield  {journal} {\bibinfo  {journal} {Phys. Rev. D}\ }\textbf {\bibinfo
  {volume} {75}},\ \bibinfo {pages} {037701} (\bibinfo {year}
  {2007})}\BibitemShut {NoStop}%
\bibitem [{\citenamefont {Patt}\ and\ \citenamefont
  {Wilczek}(2006)}]{Patt:2006fw}%
  \BibitemOpen
  \bibfield  {author} {\bibinfo {author} {\bibfnamefont {B.}~\bibnamefont
  {Patt}}\ and\ \bibinfo {author} {\bibfnamefont {F.}~\bibnamefont {Wilczek}},\
  }\href@noop {} {\  (\bibinfo {year} {2006})},\ \Eprint
  {http://arxiv.org/abs/hep-ph/0605188} {arXiv:hep-ph/0605188} \BibitemShut
  {NoStop}%
\bibitem [{\citenamefont {Krnjaic}(2016)}]{Krnjaic:2015mbs}%
  \BibitemOpen
  \bibfield  {author} {\bibinfo {author} {\bibfnamefont {G.}~\bibnamefont
  {Krnjaic}},\ }\href {\doibase 10.1103/PhysRevD.94.073009} {\bibfield
  {journal} {\bibinfo  {journal} {Phys. Rev. D}\ }\textbf {\bibinfo {volume}
  {94}},\ \bibinfo {pages} {073009} (\bibinfo {year} {2016})}\BibitemShut
  {NoStop}%
\bibitem [{\citenamefont {Winkler}(2019)}]{Winkler:2018qyg}%
  \BibitemOpen
  \bibfield  {author} {\bibinfo {author} {\bibfnamefont {M.~W.}\ \bibnamefont
  {Winkler}},\ }\href {\doibase 10.1103/PhysRevD.99.015018} {\bibfield
  {journal} {\bibinfo  {journal} {Phys. Rev. D}\ }\textbf {\bibinfo {volume}
  {99}},\ \bibinfo {pages} {015018} (\bibinfo {year} {2019})}\BibitemShut
  {NoStop}%
\bibitem [{\citenamefont {Filimonova}\ \emph {et~al.}(2020)\citenamefont
  {Filimonova}, \citenamefont {Sch\"afer},\ and\ \citenamefont
  {Westhoff}}]{Filimonova:2019tuy}%
  \BibitemOpen
  \bibfield  {author} {\bibinfo {author} {\bibfnamefont {A.}~\bibnamefont
  {Filimonova}}, \bibinfo {author} {\bibfnamefont {R.}~\bibnamefont
  {Sch\"afer}}, \ and\ \bibinfo {author} {\bibfnamefont {S.}~\bibnamefont
  {Westhoff}},\ }\href {\doibase 10.1103/PhysRevD.101.095006} {\bibfield
  {journal} {\bibinfo  {journal} {Phys. Rev. D}\ }\textbf {\bibinfo {volume}
  {101}},\ \bibinfo {pages} {095006} (\bibinfo {year} {2020})}\BibitemShut
  {NoStop}%
\bibitem [{\citenamefont {Kachanovich}\ \emph {et~al.}(2020)\citenamefont
  {Kachanovich}, \citenamefont {Nierste},\ and\ \citenamefont
  {Ni\v{s}and\v{z}i\'c}}]{Kachanovich:2020yhi}%
  \BibitemOpen
  \bibfield  {author} {\bibinfo {author} {\bibfnamefont {A.}~\bibnamefont
  {Kachanovich}}, \bibinfo {author} {\bibfnamefont {U.}~\bibnamefont
  {Nierste}}, \ and\ \bibinfo {author} {\bibfnamefont {I.}~\bibnamefont
  {Ni\v{s}and\v{z}i\'c}},\ }\href {\doibase 10.1140/epjc/s10052-020-8240-z}
  {\bibfield  {journal} {\bibinfo  {journal} {Eur. Phys. J. C}\ }\textbf
  {\bibinfo {volume} {80}},\ \bibinfo {pages} {669} (\bibinfo {year}
  {2020})}\BibitemShut {NoStop}%
\bibitem [{\citenamefont {Kamenik}\ and\ \citenamefont
  {Smith}(2009)}]{Kamenik:2009kc}%
  \BibitemOpen
  \bibfield  {author} {\bibinfo {author} {\bibfnamefont {J.~F.}\ \bibnamefont
  {Kamenik}}\ and\ \bibinfo {author} {\bibfnamefont {C.}~\bibnamefont
  {Smith}},\ }\href {\doibase 10.1016/j.physletb.2009.09.041} {\bibfield
  {journal} {\bibinfo  {journal} {Phys. Lett. B}\ }\textbf {\bibinfo {volume}
  {680}},\ \bibinfo {pages} {471} (\bibinfo {year} {2009})}\BibitemShut
  {NoStop}%
\bibitem [{\citenamefont {Jeon}\ \emph {et~al.}(2006)\citenamefont {Jeon},
  \citenamefont {Kim}, \citenamefont {Lee},\ and\ \citenamefont
  {Yu}}]{Jeon:2006nq}%
  \BibitemOpen
  \bibfield  {author} {\bibinfo {author} {\bibfnamefont {J.~H.}\ \bibnamefont
  {Jeon}}, \bibinfo {author} {\bibfnamefont {C.~S.}\ \bibnamefont {Kim}},
  \bibinfo {author} {\bibfnamefont {J.}~\bibnamefont {Lee}}, \ and\ \bibinfo
  {author} {\bibfnamefont {C.}~\bibnamefont {Yu}},\ }\href {\doibase
  10.1016/j.physletb.2006.03.069} {\bibfield  {journal} {\bibinfo  {journal}
  {Phys. Lett. B}\ }\textbf {\bibinfo {volume} {636}},\ \bibinfo {pages} {270}
  (\bibinfo {year} {2006})}\BibitemShut {NoStop}%
\bibitem [{\citenamefont {Altmannshofer}\ \emph {et~al.}(2009)\citenamefont
  {Altmannshofer}, \citenamefont {Buras}, \citenamefont {Straub},\ and\
  \citenamefont {Wick}}]{Altmannshofer:2009ma}%
  \BibitemOpen
  \bibfield  {author} {\bibinfo {author} {\bibfnamefont {W.}~\bibnamefont
  {Altmannshofer}}, \bibinfo {author} {\bibfnamefont {A.~J.}\ \bibnamefont
  {Buras}}, \bibinfo {author} {\bibfnamefont {D.~M.}\ \bibnamefont {Straub}}, \
  and\ \bibinfo {author} {\bibfnamefont {M.}~\bibnamefont {Wick}},\ }\href
  {\doibase 10.1088/1126-6708/2009/04/022} {\bibfield  {journal} {\bibinfo
  {journal} {JHEP}\ }\textbf {\bibinfo {volume} {04}},\ \bibinfo {pages} {022}
  (\bibinfo {year} {2009})}\BibitemShut {NoStop}%
\bibitem [{\citenamefont {Bartsch}\ \emph {et~al.}(2009)\citenamefont
  {Bartsch}, \citenamefont {Beylich}, \citenamefont {Buchalla},\ and\
  \citenamefont {Gao}}]{Bartsch:2009qp}%
  \BibitemOpen
  \bibfield  {author} {\bibinfo {author} {\bibfnamefont {M.}~\bibnamefont
  {Bartsch}}, \bibinfo {author} {\bibfnamefont {M.}~\bibnamefont {Beylich}},
  \bibinfo {author} {\bibfnamefont {G.}~\bibnamefont {Buchalla}}, \ and\
  \bibinfo {author} {\bibfnamefont {D.~N.}\ \bibnamefont {Gao}},\ }\href
  {\doibase 10.1088/1126-6708/2009/11/011} {\bibfield  {journal} {\bibinfo
  {journal} {JHEP}\ }\textbf {\bibinfo {volume} {11}},\ \bibinfo {pages} {011}
  (\bibinfo {year} {2009})}\BibitemShut {NoStop}%
\bibitem [{\citenamefont {Ball}\ and\ \citenamefont
  {Zwicky}(2005)}]{Ball:2004ye}%
  \BibitemOpen
  \bibfield  {author} {\bibinfo {author} {\bibfnamefont {P.}~\bibnamefont
  {Ball}}\ and\ \bibinfo {author} {\bibfnamefont {R.}~\bibnamefont {Zwicky}},\
  }\href {\doibase 10.1103/PhysRevD.71.014015} {\bibfield  {journal} {\bibinfo
  {journal} {Phys. Rev. D}\ }\textbf {\bibinfo {volume} {71}},\ \bibinfo
  {pages} {014015} (\bibinfo {year} {2005})}\BibitemShut {NoStop}%
\bibitem [{\citenamefont {Li}\ \emph {et~al.}(2019)\citenamefont {Li},
  \citenamefont {Wang}, \citenamefont {Jiang}, \citenamefont {Tan},\ and\
  \citenamefont {Wang}}]{Li:2018hgu}%
  \BibitemOpen
  \bibfield  {author} {\bibinfo {author} {\bibfnamefont {G.}~\bibnamefont
  {Li}}, \bibinfo {author} {\bibfnamefont {T.}~\bibnamefont {Wang}}, \bibinfo
  {author} {\bibfnamefont {Y.}~\bibnamefont {Jiang}}, \bibinfo {author}
  {\bibfnamefont {X.-Z.}\ \bibnamefont {Tan}}, \ and\ \bibinfo {author}
  {\bibfnamefont {G.-L.}\ \bibnamefont {Wang}},\ }\href {\doibase
  10.1007/JHEP03(2019)028} {\bibfield  {journal} {\bibinfo  {journal} {JHEP}\
  }\textbf {\bibinfo {volume} {03}},\ \bibinfo {pages} {028} (\bibinfo {year}
  {2019})}\BibitemShut {NoStop}%
\bibitem [{\citenamefont {Li}\ \emph {et~al.}(2020)\citenamefont {Li},
  \citenamefont {Wang}, \citenamefont {Jiang}, \citenamefont {Zhang},\ and\
  \citenamefont {Wang}}]{Li:2020dpc}%
  \BibitemOpen
  \bibfield  {author} {\bibinfo {author} {\bibfnamefont {G.}~\bibnamefont
  {Li}}, \bibinfo {author} {\bibfnamefont {T.}~\bibnamefont {Wang}}, \bibinfo
  {author} {\bibfnamefont {Y.}~\bibnamefont {Jiang}}, \bibinfo {author}
  {\bibfnamefont {J.-B.}\ \bibnamefont {Zhang}}, \ and\ \bibinfo {author}
  {\bibfnamefont {G.-L.}\ \bibnamefont {Wang}},\ }\href {\doibase
  10.1103/PhysRevD.102.095019} {\bibfield  {journal} {\bibinfo  {journal}
  {Phys. Rev. D}\ }\textbf {\bibinfo {volume} {102}},\ \bibinfo {pages}
  {095019} (\bibinfo {year} {2020})}\BibitemShut {NoStop}%
\bibitem [{\citenamefont {Kim}\ and\ \citenamefont {Wang}(2004)}]{Kim:2003ny}%
  \BibitemOpen
  \bibfield  {author} {\bibinfo {author} {\bibfnamefont {C.}~\bibnamefont
  {Kim}}\ and\ \bibinfo {author} {\bibfnamefont {G.-L.}\ \bibnamefont {Wang}},\
  }\href {\doibase 10.1016/j.physletb.2006.01.053} {\bibfield  {journal}
  {\bibinfo  {journal} {Phys. Lett. B}\ }\textbf {\bibinfo {volume} {584}},\
  \bibinfo {pages} {285} (\bibinfo {year} {2004})}\BibitemShut {NoStop}%
\bibitem [{\citenamefont {Wang}(2006)}]{Wang:2005qx}%
  \BibitemOpen
  \bibfield  {author} {\bibinfo {author} {\bibfnamefont {G.-L.}\ \bibnamefont
  {Wang}},\ }\href {\doibase 10.1016/j.physletb.2005.12.005} {\bibfield
  {journal} {\bibinfo  {journal} {Phys. Lett. B}\ }\textbf {\bibinfo {volume}
  {633}},\ \bibinfo {pages} {492} (\bibinfo {year} {2006})}\BibitemShut
  {NoStop}%
\bibitem [{\citenamefont {Bharucha}\ \emph {et~al.}(2016)\citenamefont
  {Bharucha}, \citenamefont {Straub},\ and\ \citenamefont
  {Zwicky}}]{Straub:2015ica}%
  \BibitemOpen
  \bibfield  {author} {\bibinfo {author} {\bibfnamefont {A.}~\bibnamefont
  {Bharucha}}, \bibinfo {author} {\bibfnamefont {D.~M.}\ \bibnamefont
  {Straub}}, \ and\ \bibinfo {author} {\bibfnamefont {R.}~\bibnamefont
  {Zwicky}},\ }\href {\doibase 10.1007/JHEP08(2016)098} {\bibfield  {journal}
  {\bibinfo  {journal} {JHEP}\ }\textbf {\bibinfo {volume} {08}},\ \bibinfo
  {pages} {098} (\bibinfo {year} {2016})}\BibitemShut {NoStop}%
\end{thebibliography}%


%

\end{document}